\documentclass[a4paper]{article}
\usepackage{epsfig}
\usepackage{graphicx}  % standard LaTeX graphics tool
\usepackage{color}     % for color figures
\usepackage{comment}
\usepackage{color}
\usepackage{lscape}
\usepackage{amssymb}
\date{\today}
 \normalsize

\newcommand{\bmat}{\left(\begin{array}}
\newcommand{\emat}{\end{array}\right)}
\newcommand{\be}{\begin{equation}}
\newcommand{\ee}{\end{equation}}
\newcommand{\bea}{\begin{eqnarray}}
\newcommand{\eea}{\end{eqnarray}}

\def\gtwid{\mathrel{\raise.3ex\hbox{$>$\kern-.75em\lower1ex\hbox{$\sim$}}}}
\def\ltwid{\mathrel{\raise.3ex\hbox{$<$\kern-.75em\lower1ex\hbox{$\sim$}}}}
\def\gev{{\rm \, Ge\kern-0.125em V}}
\def\tev{{\rm \, Te\kern-0.125em V}}

\def    \be            {\begin{equation}}
\def    \ee            {\end{equation}}
\def    \bea           {\begin{eqnarray}}
\def    \eea           {\end{eqnarray}}

\def\a{\alpha}
\def\b{\beta}
\def\g{\gamma}
\def\d{\delta}
\def\n{\nu}

\def\m{\mu}
\def\nn{\nonumber}
\def\d{\delta}
\def\D{\Delta}
\def\s{\sigma}
\def\r{\rho}
\def\t{\theta}

\def\la{\lambda}
\def\me{\langle m\rangle_e}
\def\mee{\langle m\rangle_{ee}}
\begin{document}
\renewcommand{\thefootnote}{\fnsymbol{footnote}}

\vspace{.3cm}

\title{\Large\bf Neutrino Mixing and Leptogenesis in  $\mu$\,--\,$\tau$ Symmetry
}

\author
{ \it \bf  E. I. Lashin$^{1,2,3}$\thanks{slashin@zewailcity.edu.eg, elashin@ictp.it},   N.
Chamoun$^{4,5}$\thanks{nchamoun@th.physik.uni-bonn.de}, C. Hamzaoui$^{6}$\thanks{hamzaoui.cherif@uqam.ca}
 and
S. Nasri$^{7,8}$\thanks{snasri@uaeu.ac.ae}
\\
\small$^1$ Ain Shams University, Faculty of Science, Cairo 11566,
Egypt.\\
\small$^2$ Centre for Theoretical Physics, Zewail City of Science \&
Technology, Sheikh Zayed, 6 October City, 12588, Giza, Egypt.\\
\small$^3$ The Abdus Salam ICTP, P.O. Box 586, 34100 Trieste, Italy. \\
\small$^4$  Physics Department, HIAST, P.O.Box 31983, Damascus,
Syria.\\
 \small$^5$  Physikalisches Institut der Universit$\ddot{a}$t Bonn, Nu$\ss$alle 12, D-53115 Bonn, Germany.\\
\small$^6$ Groupe de Physique Th\'eorique des Particules,
D\'epartement des Sciences de la Terre et de L'Atmosph\`ere, \\
\small Universit\'e du Qu\'ebec \`a Montr\'eal, Case Postale 8888,  Succ. Centre-Ville,
 Montr\'eal, Qu\'ebec, Canada, H3C 3P8. \\
\small$^7$ Department of Physics, UAE University, P.O.Box 17551, Al-Ain, United Arab Emirates.\\
\small$^8$ Laboratoire de Physique Th\'eorique, ES-SENIA University, DZ-31000 Oran, Algeria.\\
}
\maketitle

\begin{center}
\small{\bf Abstract}\\[3mm]
\end{center}
We study the consequences of the $Z_2$-symmetry behind  the $\m$--$\tau$ universality in neutrino mass matrix.
We then implement this symmetry in the type-I seesaw  mechanism and show how it can accommodate all sorts of lepton mass hierarchies
and generate enough lepton asymmetry to interpret the observed baryon asymmetry in the universe. We also show how a specific form of
a high-scale perturbation is kept when translated via the seesaw into the low scale domain, where it can accommodate the neutrino mixing data.
We finally present a realization of the high scale perturbed texture through addition of matter and extra exact symmetries.
\\
{\bf Keywords}: Neutrino Physics; Flavor Symmetry; Matter-anti-matter
\\
{\bf PACS numbers}: 14.60.Pq; 11.30.Hv; 98.80.Cq
\begin{minipage}[h]{14.0cm}
\end{minipage}
\vskip 0.3cm \hrule \vskip 0.5cm
%%%%%%%%%%%%%%%%%%%%%%%%%%%%%%%%%%
\section{Introduction}
Flavor symmetry is commonly used in model building seeking to determine the nine free parameters characterizing the effective neutrino mass matrix $M_\n$, namely
the three masses ($m_1, m_2$ and $m_3$), the three mixing angles ($\t_{23}, \t_{12}$ and $\t_{13}$), the two Majorana-type phases ($\r$ and $\s$) and the Dirac-type
phase ($\d$). Incorporating family symmetry at the Lagrangian level leads generally to textures of specific forms, and one may then
 study whether or not these specific textures can accommodate the experimental data involving the above mentioned parameters (\cite{LC-zeros} and references therein). The recent observation of a non-zero value for $\t_{13}$ from the T2K\cite{t2k}, MINOS\cite{mino},
and Double Chooz\cite{CHOOZ2} experiments puts constraints on models based on flavor symmetry (see Table \ref{fits} where the most recent updated neutrino oscillation parameters are taken from \cite{fog2014}). In this regard, recent,
particularly simple, choices for discrete and continuous flavor symmetry addressing the non-vanishing $\t_{13}$ question were respectively worked out (\cite{LCN-NTB} and references therein). The $\mu$--$\tau$ symmetry \cite{MUTAU1,MUTAU2} is enjoyed by many popular mixing patterns
such as tri-bimaximal mixing (TBM) \cite{tbm}, bimaximal mixing (BM) \cite{bm}, hexagonal mixing (HM) \cite{hm} and scenarios of $A_5$ mixing \cite{a5}, and it was largely studied in the literature \cite{literature}. Any form of the neutrino mass matrix respects a $(Z_2)^2$ symmetry \cite{lavsy}, and we can define the $\mu$--$\tau$ symmetry by fixing one of the two $Z_2$'s
to express exchange between the second and third families, whereas the second $Z_2$ factor is to be determined later by data or, equivalently, by $M_\n$ parameters. The whole
$(Z_2)^2$ symmetry might turn out to be a subgroup of a larger discrete group imposed on the whole leptonic sector.   In realizing $\mu$--$\tau$ symmetry we have two choices namely ($S_-, S_+$, as explained later), and
thus we have two textures corresponding to $\mu$--$\tau$ symmetry. It is known that both of these textures lead to a vanishing $\t_{13}$ (with $S_-$ achieving this in a less natural way), and thus perturbations are needed to get remedy of this situation\cite{perturbation-mu-tau}. In \cite{LCHN1} we studied the perturbed $\m$--$\tau$ neutrino symmetry and found the four patterns, obtained by disentangling the effects of the perturbations, to be phenomenologically viable.
\begin{table}[h]
\caption{Allowed $3\s$-ranges for the neutrino oscillation parameters, mixing angles and mass-square differences, taken from the global fit to neutrino oscillation data \cite{fog2014}.  The quantities $\d m^2$ and $\Delta m^2$ are respectively defined as  $m_2^2-m_1^2$ and  $m_3^2 -\left(m_1^2 + m_2^2\right)/2$, whereas $R_\n$ denotes the phenomenologically important quantity ${\d m^2 \over \left|\Delta m^2\right|}$. Normal and Inverted Hierarchies are
respectively denoted by NH and IH.}
\centering
\begin{tabular}{lcc}
\hline
\hline
Parameter & Best fit & $3\sigma$ range  \\
\hline
$\d m^2 \left(10^{-5} \mbox{eV}^2\right)$ & $7.54$ & $6.99-8.18$ \\
$\left|\Delta m^2\right| \left(10^{-3} \mbox{eV}^2\right)$ (NH)  & $2.43$ & $2.23-2.61$ \\
$\left|\Delta m^2\right| \left(10^{-3} \mbox{eV}^2\right)$ (IH)  & $2.38$ & $2.19-2.56$ \\
$R_\n$ (NH)  & $0.0310$ & $0.0268-0.0367$ \\
$R_\n$ (IH)  & $0.0317$ & $0.0273-0.0374$ \\
$\theta_{12}$ (NH or IH)\hspace{2cm} & $33.71^0$ & $30.59^0-36.80^0$  \\
%%%%%%%%%%%%%%%%%%%%%%%%%%%%%%%%%%%%%%%%%%%%%%%%%%
$\theta_{13}$ (NH) & $8.80^0$ & $7.62^0-9.89^0$   \\
$\theta_{13}$ (IH) & $8.91^0$ & $7.67^0-9.94^0$   \\
%---------------------------------------------------------------------
$\theta_{23}$ (NH) & $ 41.38^0$ & $37.69^0-52.30^0$   \\
$\theta_{23}$ (IH) & $38.07^0$ & $38.07^0-53.19^0$   \\
\hline
\hline
\end{tabular}
\label{fits}
\end{table}

In this work, we re-examine the question of exact $\m$--$\tau$ symmetry and implement it in a complete setup of the leptonic sector.  Then, within type-I seesaw scenarios, we show
the ability of exact symmetry to accommodate lepton mass hierarchies. Upon
studying its effect on leptogenesis we find, in contrast to other symmetries studied in \cite{LCN-NTB} and \cite{LCN-z23u1} that it can account for it. The reason behind this fact is
that fixing just one $Z_2$ in $\mu$--$\tau$ symmetry leaves one mixing angle free which can be adjusted differently in the Majorana and Dirac neutrino mass matrices ($M_R$ and $M_D$),
thus allowing for different diagonalizing matrices. For the mixing angles and in order to accommodate data, we introduce perturbations at the seesaw high scale and study their propagations
 into the low scale effective neutrino mass matrix. As in \cite{LCHN1}, we consider that the perturbed texture arising at the high scale keeps its form upon
 RG running which, in accordance with \cite{RGrunning}, does not affect the results in many setups. As to the origin of the perturbations, we shall not introduce explicitly
symmetry breaking terms into the Lagrangian \cite{Ross-Hall}, but rather follow \cite{LCHN1}, and enlarge the symmetry with extra matter and then spontaneously
break the symmetry by giving vacuum expectation values (vev) to the involved Higgs fields.

The plan of the paper is as follows. In Section 2, we review the standard notation for the neutrino mass matrix and the definition of the $\m$--$\tau$ symmetry. In Section 3 and 4,
 we introduce the two textures realizing the $\m$--$\tau$ symmetry through $S_-$ and $S_+$ respectively. We then specify our analysis to the latter case ($S_+$), and in Section 5 we introduce  the type-I seesaw scenario. We address the charged lepton sector in Subsection 5.1, whereas we study the different neutrino mass hierarchies in Subsection 5.2, and in Subsection 5.3, we study the generation of lepton asymmetry. Sections 6 and 7 examine the possible consequences for one particular possible deviation from the exact $\m$--$\tau$ symmetry, where
  we present the analytical study in the former section, while the numerical study is given in the latter section. In Section 8 we present
a theoretical realization of the perturbed texture. We end by discussion and summary in Section 9.

\section{Notations and  preliminaries}
In the Standard
Model (SM) of particle interactions, there are 3 lepton families.
The charged-lepton mass matrix linking left-handed (LH) to their
right-handed (RH) counterparts is arbitrary, but can always be
diagonalized by a bi-unitary transformation:
\begin{equation}
V^l_L\; {M}_l\; (V^l_R)^\dagger =  \pmatrix{m_e & 0 & 0 \cr 0 & m_\mu & 0 \cr 0 & 0 &
m_\tau} .
\end{equation}
Likewise, we can diagonalize the symmetric Majorana neutrino mass matrix by just one unitary transformation:
\be
V^{\n \dagger} M_\nu\; V^{\n *} \; = \; \left (\matrix{ m_1 & 0 & 0 \cr 0 & m_2 & 0
\cr 0 & 0 & m_3 \cr} \right ), \;
\label{diagM}
\ee
with $m_i$ (for $i=1,2,3$) real and positive.

The
observed neutrino mixing matrix comes from the mismatch between
$V^l$ and $V^\nu$ in that
\begin{eqnarray}
V_{\mbox{\tiny PMNS}} &=& (V^l_L)^\dagger\; V^\nu  .
\end{eqnarray}
If the charged lepton mass eigen states are the same as the current (gauge) eigen states, then $V^l_L ={\bf 1}$ (the unity matrix) and the
measured mixing comes only from the neutrinos $V_{\mbox{\tiny PMNS}} = V^{\nu}$. We shall assume this saying that we
 are working in the ``flavor" basis. As we shall see, corrections due to $V^l_L \neq {\bf 1}$ are expected to be of order of ratios of
 the hierarchical
 charged lepton masses, which are small enough to justify our assumption of working in the flavor basis. However, one can treat these corrections
 as small perturbations and embark on a phenomenological analysis involving them \cite{Ross-Hall}.

We shall adopt the parametrization of \cite{Xing}, related to other ones by simple relations \cite{LC-zeros}, where the $V_{\mbox{\tiny PMNS}}$
is given in terms of three mixing
angles $(\theta_{12}, \theta_{23}, \theta_{13})$ and three phases ($\delta,\rho,\sigma$),  as follows.
\bea
P &=& \mbox{diag}\left(e^{i\rho},e^{i\sigma},1\right)\,, \nn\\
U \; &=& \nn R_{23}\left(\t_{23}\right)\; R_{13}\left(\t_{13}\right)\; \mbox{diag}\left(1,e^{-i\d},1\right)\; R_{12}\left(\t_{12}\right)\, \\ &=& \;
\left ( \matrix{ c_{12}\, c_{13} & s_{12}\, c_{13} & s_{13} \cr - c_{12}\, s_{23}
\,s_{13} - s_{12}\, c_{23}\, e^{-i\delta} & - s_{12}\, s_{23}\, s_{13} + c_{12}\, c_{23}\, e^{-i\delta}
& s_{23}\, c_{13}\, \cr - c_{12}\, c_{23}\, s_{13} + s_{12}\, s_{23}\, e^{-i\delta} & - s_{12}\, c_{23}\, s_{13}
- c_{12}\, s_{23}\, e^{-i\delta} & c_{23}\, c_{13} \cr } \right ) \; ,\nn\\
V_{\mbox{\tiny PMNS}} &=& U\;P\, = \left ( \matrix{ c_{12}\, c_{13} e^{i\rho} & s_{12}\, c_{13} e^{i\sigma}& s_{13} \cr (- c_{12}\, s_{23}
\,s_{13} - s_{12}\, c_{23}\, e^{-i\delta}) e^{i\rho} & (- s_{12}\, s_{23}\, s_{13} + c_{12}\, c_{23}\, e^{-i\delta})e^{i\sigma}
& s_{23}\, c_{13}\, \cr (- c_{12}\, c_{23}\, s_{13} + s_{12}\, s_{23}\, e^{-i\delta})e^{i\rho} & (- s_{12}\, c_{23}\, s_{13}
- c_{12}\, s_{23}\, e^{-i\delta})e^{i\sigma} & c_{23}\, c_{13} \cr } \right ),
\label{defv}
\eea
where $R_{ij}\left(\t_{ij}\right)$ is the rotation matrix in the $(i,j)$-plane by angle $\t_{ij}$, and $s_{12} \equiv \sin\theta_{12} \ldots$. Note that in this adopted parametrization, the
third column of $V_{\mbox{\tiny PMNS}}$ is real.

In this parametrization, and in the flavor basis, the neutrino mass matrix elements
are given by:
\bea
M_{\n\,11}&=& m_1\, c_{12}^2\, c_{13}^2\, e^{2\,i\,\r} + m_2\, s_{12}^2\, c_{13}^2\, e^{2\,i\,\s}
+ m_3\,s_{13}^2,\nn\\
%%%%%%%%%%%%%%%%%%%%%%%%%%%%%%%%%%%%%
M_{\n\,12}&=& m_1\,\left( - c_{13}\, s_{13}\, c_{12}^2 \,s_{23} e^{2\,i\,\r}
- c_{13}\, c_{12} s_{12}\, c_{23}\, e^{i\,(2\,\r-\d)}\right)\nn \\ &&
+ m_2\,\left( - c_{13}\, s_{13}\, s_{12}^2\, s_{23} e^{2\,i\,\s}
+ c_{13}\, c_{12}\, s_{12}\, c_{23}\, e^{i\,(2\,\s-\d)}\right) + m_3\, c_{13}\, s_{13}\, s_{23},\nn\\
%%%%%%%%%%%%%%%%%%%%%%%%%%%%%%%%%%
M_{\n\,13}&=& m_1\,\left( - c_{13}\, s_{13}\, c_{12}^2\, c_{23}\, e^{2\,i\,\r}
+ c_{13}\, c_{12}\, s_{12}\, s_{23}\, e^{i\,(2\,\r-\d)}\right)\nn \\ &&
+ m_2\,\left( - c_{13}\, s_{13}\, s_{12}^2\, c_{23}\, e^{2\,i\,\s}
- c_{13}\, c_{12}\, s_{12}\, s_{23}\, e^{i\,(2\,\s-\d)}\right) + m_3\, c_{13}\, s_{13}\, c_{23},\nn\\
%%%%%%%%%%%%%%%%%%%%%%%%%%%%%%%%%
M_{\n\,22}&=& m_1\, \left( c_{12}\, s_{13}\, s_{23}\,  e^{i\,\r}
+ c_{23}\, s_{12}\, e^{i\,(\r-\d)}\right)^2\nn \\ &&
 + m_2\, \left( s_{12}\, s_{13}\, s_{23}\,  e^{i\,\s}
- c_{23}\, c_{12}\, e^{i\,(\s-\d)}\right)^2 + m_3\, c_{13}^2\, s_{23}^2, \nn\\
%%%%%%%%%%%%%%%%%%%%%%%%%%%%%%%%%%%%%%%%%%%%
M_{\n\,33}&=& m_1\, \left( c_{12}\, s_{13}\, c_{23}\,  e^{i\,\r}
- s_{23}\, s_{12}\, e^{i\,(\r-\d)}\right)^2\nn \\ &&
 + m_2\, \left( s_{12}\, s_{13}\, c_{23}\,  e^{i\,\s}
+ s_{23}\, c_{12}\, e^{i\,(\s-\d)}\right)^2 + m_3\, c_{13}^2\, c_{23}^2, \nn\\
%%%%%%%%%%%%%%%%%%%%%%%%%%
M_{\n\, 23} &=& m_1\,\left( c_{12}^2\, c_{23}\, s_{23}\, s_{13}^2\,  e^{2\,i\,\r}
 + s_{13}\, c_{12}\, s_{12}\, \left(c_{23}^2-s_{23}^2\right)\, e^{i\,(2\,\r-\d)} - c_{23}\, s_{23}\, s_{12}^2\, e^{2\,i\,(\r-\d)}\right)
\nn\\
&&
+  m_2\,\left( s_{12}^2\, c_{23}\, s_{23}\, s_{13}^2\,  e^{2\,i\,\s}
 + s_{13}\, c_{12}\, s_{12}\, \left(s_{23}^2-c_{23}^2\right)\, e^{i\,(2\,\s-\d)} - c_{23}\, s_{23}\, c_{12}^2\, e^{2\,i\,(\s-\d)}\right)\nn \\ &&
+ m_3\, s_{23}\, c_{23}\, c_{13}^2.
\label{melements}
\eea
This helps in viewing directly at the level of the mass matrix that the effect of swapping the indices $2$ and $3$ corresponds to the transformation $\t_{23} \rightarrow \frac{\pi}{2} -
\t_{23}$ and $\d \rightarrow \d \pm \pi$. Hence, for a texture satisfying the $\mu$-$\tau$ symmetry, one can check the correctness of any obtained formula by requesting it to be invariant under the above transformation.

As said before, any form of $M_\n$ satisfies a $Z_2^2$-symmetry. This means that there are two commuting unitary $Z_2$-matrices (squared to unity) ($S_1, S_2$) which leave  $M_\n$ invariant:
\begin{eqnarray}\label{FI}
S^{T}\;M_{\nu}\;S = M_{\nu}
\end{eqnarray}
For a non-degenrate mass spectrum, the form of the $Z_2$-matrix $S$ is given by \cite{LCN-z23u1}:
\bea
S &=& {V^\n} \;\mbox{diag}(\pm 1, \pm 1 , \pm 1)\; {V^\n}^{\dagger}
\eea
where the two $S$'s correspond to having, in $\mbox{diag}(\pm 1, \pm 1 , \pm 1)$, two pluses and one minus, the position of which differs in the two $S$'s (the third $Z_2$-matrix, corresponding to the third position of the minus sign, is generated by multiplying the two $S$'s and noting that the form invariance formula Eq.(\ref{FI}) is invariant under $S \rightarrow -S$).

In practice, however, we follow a reversed path, in that if we assume a `real' orthogonal $Z_2$-matrix (and hence symmetric with eigenvalues $\pm1$) satisfying Eq.(\ref{FI}), then it commutes
 with $M_\n$, and so both matrices can be simultaneously diagonalized. Quite often, the form of $S$ is simpler than $M_\n$, so one proceeds to solve the eigensystem problem for $S$, and find a unitary diagonalizing  matrix $\tilde{U}$:
 \bea \tilde{U}^\dagger \; S\; \tilde{U} &=& \mbox{Diag}\left(\pm 1,\pm 1, \pm 1\right) \label{diagS}
 \eea
 The conjugate matrix $\tilde{U}^*$ can `commonly' be identified with, or related simply to, the matrix $V$ satisfying Eq.(\ref{diagM})\footnote{In fact, as we shall see, starting from the
 general form of $\tilde{U}$ satisfying Eq.(\ref{diagS}), one can determine (up to a diagonal phase matrix) the unitary matrix $\tilde{U}_0$ which diagonalizes
  simultaneously the two commuting hermitian matrices $S$ and $M^*_\nu M_\nu$ so that $\tilde{U}_0^\dagger \; M^*_\nu M_\nu\; \tilde{U}_0 = \mbox{Diag}\left(m_1^2,m_2^2, m_3^2\right)
  =D^2$. One can show then that $D^2$ commutes with $\tilde{U}_0^T M_\nu \tilde{U}_0$  which leads to the latter matrix being diagonal. Fixing now the phases so that the latter diagonal matrix becomes real  makes $\tilde{U}_0$ play the role of $V^*$ in Eq. (\ref{diagM}). One then can use the freedom in rephasing the charged lepton fields to force the adopted parametrization on $V_{\mbox{\tiny PMNS}}$.}. In this case, and in the flavor basis, the $V_{\mbox{\tiny PMNS}}$ would be generally complex  and equal to the one presented in Eq.(\ref{defv}). Determining the eigenvectors of the $S$ matrices helps thus to determine the neutrino mixing and phase angles.

The $\m$--$\tau$ symmetry is defined when one of the two $Z_2$-matrices corresponds to switching between the $2^{\mbox{nd}}$ and the $3^{\mbox{rd}}$ families. We have, up to a global irrelevant minus sign (see again Eq.\ref{FI}), two choices, which would lead to two textures at the level of $M_\n$.

\section{\large The $\mu$--$\tau$ symmetry manifested through $S_{-}$: $\left(M_{\nu\,12}=M_{\nu\,13}\; \mbox{and} \; M_{\nu\,22}=M_{\nu\,33}\right)$}
The $Z_2$-symmetry matrix is given by:
\begin{equation}
S_{-} =\left(\begin{array}{ccc} 1 & 0 & 0 \\
0 & 0 & 1 \\
0 & 1 & 0 \end{array}\right)
\end{equation}
The invariance of $M_\n$ under $S_-$ (Eq.\ref{FI})
forces the symmetric matrix $M_\n$ to have a texture of the form:
%%%%%%%%%%%%%%%%%%%%
\bea
M_{\nu} &=& \left(\begin{array}{ccc}
A_{\nu} & B_{\nu} &  B_{\nu} \\
B_{\nu} & C_{\nu}  & D_{\nu} \\
B_{\nu} & D_{\nu} & C_{\nu} \end{array}\right).
\label{massmatrix-}
\eea
%%%%%%%%%%%%%%%%%%%%%%

The invariance of $M_{\nu}$ under $ S_{-}$ implies that $ S_{-}$ commutes with both $M_{\nu}$ and  $M_{\nu}^*$, and thus also with the hermitian positive matrices $M_{\nu}^* M_{\nu}$ and $M_{\nu} M_{\nu}^*$. One can easily find the general form of the diagonalizing unitary matrix of $S_-$ (up to an arbitrary diagonal phase matrix). The matrix $S_-$ has normalized eigen vectors:
$\left\{ v_1= \left(0, 1/\sqrt{2}, 1/\sqrt{2}\,\right)^{\mbox{\tiny T}}, \, v_2= \left(1,0,0\,\right)^{\mbox{\tiny T}}, \, v_3=\left(0, 1/\sqrt{2} , -1/\sqrt{2}\,\right)^{\mbox{\tiny T}} \right\}$ corresponding respectively to the
eigenvalues $\left( 1, 1, -1\right)$. Since the eigenvalue $1$ is two-fold degenerate, then there is still freedom for a unitary transformation defined by an angle $\varphi$ and phase
$\xi$ in its eigenspace to get the new eigen vectors in the following form:
\bea
\label{eigenvectorrotation}
\overline{v}_1 & =& - s_\varphi\, e^{i\,\xi}\, v_1 +  c_\varphi\, v_2,\nn \\
\overline{v}_2 & = & c_\varphi\, e^{i\,\xi}\, v_1 + s_\varphi\, v_2,
\eea
We have three choices as to how we order the eigenvectors forming the diagonalizing matrix $U$, and we chose the one which would lead to ``plausible" mixing angles falling in the first quadrant.
This choice for ordering the eigenvalues turns out to be $(1,-1,1)$, as we could check that the two choices corresponding to the other two positions for the eigenvalue ($-1$) lead upon identification with $V_{\mbox{\tiny PMNS}}$ in Eq.(\ref{defv}) to some mixing angles lying outside the first quadrant, and the matrix $U_-$ which diagonalizes $S_-$ can be cast into the form:
\bea \label{U-}
U_{-} = \left[\overline{v}_1, v_3, \overline{v}_2\right] =  \left(\begin{array}{ccc}
 c_ \varphi & 0 &  s_ \varphi \\
- s_\varphi\,e^{i\,\xi} /\sqrt{2}  & 1/\sqrt{2} & c_\varphi\, e^{i\,\xi}/\sqrt{2}\\
 - s_\varphi\,e^{i\,\xi}/\sqrt{2} & -1/\sqrt{2} & c_\varphi\, e^{i\,\xi}/\sqrt{2} \end{array}\right).
\eea

One can single out of this general form the unitary matrix which diagonalizes also the hermitian positive matrix $M_{\nu}^* \,M_{\nu}$ with different positive eigenvalues.
In order to simplify the resulting formulas, the matrix $M_{\nu}^* \,M_{\nu}$ can be organized in a concise form as,
\bea \label{M*M}
 M_{\nu}^* \,M_{\nu} &= &
 \left(\begin{array}{ccc}
 a_\n & b_\n &  b_\n \\
b_\n^* &  c_\n & d_\n\\
b_\n^* & d_\n & c_\n
\end{array}\right),
\eea
where  $a_\n$ , $b_\n$, $c_\n$ and $d_\n$ are defined as follows,
\bea
  a_\n  =  \left|A_\n\right|^2 + 2 \left|B_\n\right|^2, && b_\n  = A_\n^*\,B_\n + B_\n^*\,C_\n + B_\n^*\,D_\n,\nn \\
  c_\n  =  \left|A_\n\right|^2 +  \left|B_\n\right|^2 + \left|C_\n\right|^2, &&
  d_\n  =  \left|B_\n\right|^2 + C_\n^*\,D_\n + D_\n^*\,C_\n.
  \eea
The diagonalization of $M_{\nu}^* \,M_{\nu}$ through $U_-$ fixes $\varphi$ and $\xi$ to be:
\bea
\label{varphi} \tan \left(2 \varphi\right) =
\frac{2 \, \sqrt{2} \, \left|b_\n\right| }
{c_{\nu} + d_{\nu} - a_{\nu} }, && \xi = \mbox{Arg}\left(b_\n^*\right).
\eea
Now and after having fixed  $\varphi$ and $\xi$ we have,
\be
U_-^{\dagger}\,M_{\nu}^* \,M_{\nu} U_- = U_-^{T}\,M_{\nu} \,M_{\nu}^*\, U_-^*  = \mbox{Diag}\left(m_1^2,m_2^2, m_3^2\right),
\label{diagm1}
\ee
where \bea
m_1^2 & = & {a_\n + c_\n + d_\n\over 2} + {1\over 2} \sqrt{\left(a_\n-d_\n-c_\n\right)^2 + 8\, \left|b_\n\right|^2},\nn\\
 m_2^2  & = & c_\n - d_\n,\nn\\
m_3^2 & = & {a_\n + c_\n + d_\n\over 2} - {1\over 2} \sqrt{\left(a_\n-d_\n-c_\n\right)^2 + 8\, \left|b_\n\right|^2}.
\eea

The above relations imply directly that $U_-^{T}\,M_{\nu}\, U_-$ commutes with $(U_-^{T}\,M_{\nu}\, U_-)^*$, and hence also with the product of these two matrices which is
a diagonal matrix: $U_-^{T}\,M_{\nu}\, U_-(U_-^{T}\,M_{\nu}\, U_-)^*=U_-^{T}\,M_{\nu} \,M_{\nu}^*\, U_-^* $. Since we have a non-degenrate spectrum amounting
to
different eigenvalues of $M_{\nu} \,M_{\nu}^*$, we deduce directly that  $U_-^{T}\,M_{\nu}\, U_-$ is diagonal. Actually we get:
\be
 U_-^T\,M_{\nu}\, U_- = M_\n^{\mbox{\tiny Diag}},
\ee
where $M_\n^{\mbox{\tiny Diag}}$ is a diagonal matrix whose entries are,
\bea
M_{\n\,11}^{\mbox{\tiny Diag}} &=& A_\n\,c_\varphi^2 - \sqrt{2}\,s_{2\varphi}\,e^{i\,\xi}\,B_\n + \left(C_\n + D_\n\right)\,s_\varphi^2\,e^{2\,i\,\xi},\nn\\
M_{\n\,22}^{\mbox{\tiny Diag}} &=& C_\n - D_\n,\nn\\
M_{\n\,33}^{\mbox{\tiny Diag}} &=& A_\n\,s_\varphi^2 + \sqrt{2}\,s_{2\varphi}\,e^{i\,\xi}\,B_\n + \left(C_\n + D_\n\right)\,c_\varphi^2\,e^{2\,i\,\xi}.
\eea

In order to extract the mixing and phase angles, we use the freedom of multiplying $U_-$ by a diagonal phase matrix $Q=\mbox{Diag}\left(e^{-ip_1},e^{-ip_2}, e^{-ip_3}\right)$ to ensure real positive eigenvalues for the mass matrix $M_\n$ such that
\bea (U_- \;\;Q)^T M_\nu (U_- \;\;Q) &=& \mbox{Diag}\left(m_1,m_2, m_3\right),
\label{U_-Q}
\eea
and we find that we should take
\be p_i = \frac{1}{2}\, \mbox{Arg}(M_{\n_{ii}}^{\mbox{\tiny Diag}}),\;\;i=1,2,3.
\ee

However, we get now the following form for the diagonalizing matrix $U_-\;\;Q$:
\bea
U_-\;\;Q &=& \left ( \matrix{ c_{\phi}\,e^{-ip_1} & 0 & s_{\phi}e^{-ip_3} \cr -\frac{1}{\sqrt{2}} s_\phi e^{i (\xi - p_1)} & \frac{1}{\sqrt{2}}e^{-ip_2}
& \frac{1}{\sqrt{2}} c_\phi e^{i (\xi - p_3)}  \cr -\frac{1}{\sqrt{2}} s_\phi e^{i (\xi - p_1)} & -\frac{1}{\sqrt{2}}e^{-ip_2} & \frac{1}{\sqrt{2}} c_\phi e^{i (\xi - p_3)} \cr } \right ),
\eea
In order to have the conjugate of this matrix in the same form as the adopted parametrization of $V_{\mbox{\tiny PMNS}}$ in Eq.(\ref{defv}), where the third column is real, we can make a phase change in the
charged lepton fields:
\bea
e\rightarrow e^{-i p_3}\; e,\; \mu \rightarrow e^{i (\xi - p_3)}\; \mu ,\; \tau \rightarrow e^{i (\xi - p_3)}\; \tau
\eea
so that we identify now the mixing and phase angles and see that
 the $\mu$--$\tau$ symmetry forces the following angles:
 \bea
 \label{mixingS-}
 &&\t_{23}= \pi / 4,\;\; \t_{12}=0 ,\;\; \t_{13}= \varphi,\nn \\
 && \rho = {1\over 2}\, \mbox{Arg}\left(M_{\n\,11}^{\mbox{\tiny Diag}}\,M_{\n\,33}^{\mbox{\tiny Diag\,*}}\right),\;\;
 \sigma = {1\over 2}\,\mbox{Arg}\left(M_{\n\,22}^{\mbox{\tiny Diag}}\,M_{\n\,33}^{\mbox{\tiny Diag\,*}}\right),\;\; \delta = 2\,\pi - \xi.
 \eea
We can get, as phenomenology suggests, a small value for $\t_{13}$  assuming
\bea
\label{condtune}
\left|b_ \nu\right| &\ll& \left|c_\n+d_\n - a_\n\right|.
\eea
and then the mass spectrum turns out to be:
\bea
\label{massspecS-}
%%%%%%%%%%%%%%%%%%%%%%%
m_1^2 \approx a_\n, & m_2^2 = c_\n - d_\n,& m_3^2 \approx c_\n + d_\n
\eea
Inverting these relations to express the mass parameters in terms of the mass eigenvalues we get these simple direct relations,
\bea
a_\n \approx m_1^2, & c_\n \approx {m_2^2 + m_3^2\over 2}, &
d_\n \approx {m_3^2 - m_2^2\over 2}.
\label{relm1}
\eea
It is remarkable that all kinds of  mass spectra can be accommodated by properly adjusting the parameters $a_\n, c_\n, $ and $d_\n$ according to the relations in Eq.(\ref{relm1}).
As to the mixing angles, we see that the value of $\t_{23}$ is phenomenologically acceptable corresponding to maximal atmospheric  mixing, and the parameter $b_\n$ can be adjusted according to
Eq.(\ref{condtune}) to accommodate the small mixing angle $\t_{13}$. The phases are not of much concern because so far there is no serious constraint on phases.
It seems that all things fit properly except the vanishing value of the mixing angle $\t_{12}$ which is far from its experimental value  $\simeq 33.7^o$.

One might argue that this symmetry pattern $S_-$  might be viable phenomenologically if we adopt an alternative choice of ordering its eigenvalues and use the phase ambiguity to put all mixing angles in the first quadrant. We have not done this, but rather we prefer to find a phenomenologically viable symmetry leading directly to mixing angles in the first quadrant. This can be
carried out in the second texture expressing the $\m$--$\tau$ symmetry materialized through $S_+$.

\section{\large The $\mu$--$\tau$ symmetry manifested through $S_{+}$: $\left(M_{\nu\,12}=-M_{\nu\,13}\; \mbox{and} \; M_{\nu\,22}=M_{\nu\,33}\right)$}
The $Z_2$-symmetry matrix is given by:
\begin{equation}
S_{+} =\left(\begin{array}{ccc} -1 & 0 & 0 \\
0 & 0 & 1 \\
0 & 1 & 0 \end{array}\right)
\end{equation}
The invariance of $M_\n$ under $S_+$ (Eq. \ref{FI})
forces the symmetric matrix $M_\n$ to have a texture of the form:
%%%%%%%%%%%%%%%%%%%%
\bea
M_{\nu} &=& \left(\begin{array}{ccc}
A_{\nu} & B_{\nu} &  -B_{\nu} \\
B_{\nu} & C_{\nu}  & D_{\nu} \\
-B_{\nu} & D_{\nu} & C_{\nu} \end{array}\right) \label{massmatrix+}
\eea
%%%%%%%%%%%%%%%%%%%%%%

As before, $ S_{+}$ commutes with  $M_{\nu}$, $M_{\nu}^*$ and thus also with $M_{\nu}^* M_{\nu}$ and $M_{\nu} M_{\nu}^*$. The normalized eigen vectors of $ S_{+}$ are:
$\left\{ v_1= \left(\,0, -1/\sqrt{2}, 1/\sqrt{2}\,\right)^{\mbox{\tiny T}}, \, v_2= \left(\,1,0,0\,\right)^{\mbox{\tiny T}}, \, v_3=\left(\,0, 1/\sqrt{2} , 1/\sqrt{2}\,\right)^{\mbox{\tiny T}} \right\}$ corresponding respectively to the
eigenvalues $\left\{- 1,- 1, 1\right\}$. We would like to find the general form (up to a diagonal phase matrix) of the unitary diagonalizing
matrix of $S_+$. Since the eigenvalue $-1$ is two-fold degenerate, then there is still freedom for a unitary transformation defined  by an angle $\varphi$ and phase $\xi$ in its eigenspace to get new eigen vectors in the following form:
\bea
\label{eigenvecs+}
\overline{v}_1 & =& s_\varphi\, e^{-i\,\xi}\, v_1 +  c_\varphi\, v_2,\nn \\
\overline{v}_2 & = & -c_\varphi\, e^{-i\,\xi}\, v_1 + s_\varphi\, v_2.
\eea
Once again, the suitable choice of ordering the eigenvectors of $S_+$, which would determine the unitary matrix $U_{+}$ diagonalizing $S_{+}$ in such a way that the
mixing angles fall all in the first quadrant, turns out to correspond to the eigenvalues ordering $\left\{- 1,- 1, 1\right\}$. Hence, the matrix $U_{+}$
 assumes  the following form:
 \bea
 \label{U+}
U_{+} = \left[\overline{v}_1, \overline{v}_2, v_3\right] =  \left(\begin{array}{ccc}
 c_\varphi & s_\varphi &  0 \\
- s_\varphi\,e^{-i\,\xi} /\sqrt{2}  & 1/\sqrt{2}\,c_\varphi\,e^{-i\,\xi} & 1/\sqrt{2}\\
  s_\varphi\,e^{-i\,\xi}/\sqrt{2} & -1/\sqrt{2}\,c_\varphi\,e^{-i\,\xi} & 1/\sqrt{2} \end{array}\right).
\eea
The matrix $M_{\nu}^* M_{\nu}$ has the form,
\bea \label{M*MS+}
 M_{\nu}^* \,M_{\nu} &= &
 \left(\begin{array}{ccc}
 a_\n & b_\n &  b_\n \\
b_\n^* &  c_\n & d_\n\\
-b_\n^* & d_\n & c_\n
\end{array}\right),
\eea
where  $a_\n$ , $b_\n$, $c_\n$ and $d_\n$ are defined as follows,
\bea
  a_\n  =  \left|A_\n\right|^2 + 2 \left|B_\n\right|^2, && b_\n  = A_\n^*\,B_\n + B_\n^*\,C_\n - B_\n^*\,D_\n,\nn \\
  c_\n  =  \left|B_\n\right|^2 +  \left|C_\n\right|^2 + \left|D_\n\right|^2, &&
  d_\n  =  -\left|B_\n\right|^2 + C_\n^*\,D_\n + D_\n^*\,C_\n.
  \label{abs+}
  \eea and its eigenvalues are given by:
  \bea
m_1^2 & = & {a_\n + c_\n - d_\n\over 2} + {1\over 2} \sqrt{\left(a_\n+d_\n-c_\n\right)^2 + 8\, \left|b_\n\right|^2},\nn\\
 m_2^2  & = & {a_\n + c_\n - d_\n\over 2} - {1\over 2} \sqrt{\left(a_\n+d_\n-c_\n\right)^2 + 8\, \left|b_\n\right|^2},\nn\\
m_3^2 & = & c_\n + d_\n.
\label{msqs+}
\eea
The specific form of $U_+$ of Eq.(\ref{U+}) which diagonlizes also the hermitian matrix $M_\n^* M_\n$, which commutes with $S_+$, corresponds to:
\bea
\label{varphi} \tan \left(2 \varphi\right) =
\frac{2 \, \sqrt{2} \, \left|b_\n\right| }
{c_{\nu} - a_{\nu} - d_{\nu} }, && \xi = \mbox{Arg}\left(b_\n\right),
\label{vars+}
\eea
As in the case of $U_-$, one can prove that $U_+^{T}\; M_\n \;U_+$, after having fixed $\varphi$ and $\xi$ according to Eq. (\ref{varphi}), is diagonal
 \be
 U_+^{T}\;M_{\nu}\; U_+ = M_\n^{\mbox{\tiny Diag}}= {\mbox{Diag} \left( M_{\n\,11}^{\mbox{\tiny Diag}},\; M_{\n\,22}^{\mbox{\tiny Diag}},\; M_{\n\,33}^{\mbox{\tiny Diag}} \right)} ,
\label{u+}
\ee
where
\bea
M_{\n\,11}^{\mbox{\tiny Diag}} &=& A_\n\,c_\varphi^2 - \sqrt{2}\,s_{2\varphi}\,e^{-i\,\xi}\,B_\n + \left(C_\n - D_\n\right)\,s_\varphi^2\,e^{-2\,i\,\xi},\nn\\
M_{\n\,22}^{\mbox{\tiny Diag}} &=& A_\n\,s_\varphi^2 + \sqrt{2}\,s_{2\varphi}\,e^{-i\,\xi}\,B_\n + \left(C_\n - D_\n\right)\,c_\varphi^2\,e^{-2\,i\,\xi},\nn\\
M_{\n\,33}^{\mbox{\tiny Diag}} &=& C_\n + D_\n,
\label{Mdiags+}
\eea
while the squared modulus of these complex eigenvalues are identified respectively with the squared mass $m_1^2$, $m_2^2$ and $m_3^2$ (the eigenvalues of $M_{\nu}^* M_{\nu}$ in Eq. \ref{msqs+} ).

Again, as was the case for the $S_-$ pattern, we use the freedom of multiplying $U_+$ by a diagonal phase matrix $Q$ in order that
\bea
(U_+\; Q)^T M_\nu (U_+\; Q) &=& \mbox{Diag}\left(m_1,\;m_2,\; m_3\right).
\label{U_+Q}
\eea
Moreover, we re-phase the charged lepton fields to make the conjugate of $(U_+\; Q)$ in the same form
as the adopted parametrization for $V_{\mbox{\tiny{PMNS}}}$ in Eq.(\ref{defv}), so that to identify the mixing and phase angles. We find that the $\mu$--$\tau$ symmetry realized through $S_+$ entails the followings:
 \bea
 \label{mixingS+}
 &&\t_{23}= \pi / 4,\;\; \t_{12}=\varphi ,\;\; \t_{13}= 0,\nn \\
 && \rho = {1\over 2}\, \mbox{Arg}\left(M_{\n\,11}^{\mbox{\tiny Diag}}\right),\;\;
 \sigma = {1\over 2}\,\mbox{Arg}\left(M_{\n\,22}^{\mbox{\tiny Diag}}\right),\;\; \delta = {1\over 2}\,\mbox{Arg}\left(M_{\n\,33}^{\mbox{\tiny Diag}}\right) - \xi.
 \eea
These predictions are phenomenologically ``almost" viable (the non-vanishing value of $\t_{13}$ will be attributed to small deviations from the exact symmetry), and furthermore do not require a special adjustment for the parameters $a_\n, b_\n, c_\n, d_\n$ which can be
of the same order, in contrast to Eq.(\ref{condtune}), and still accommodate the experimental value of $\t_{12} \simeq 33.7^o$.

The various neutrino mass hierarchies can also be produced as can be seen from Eq.(\ref{msqs+}) and Eq.(\ref{vars+}) where the three masses and the angle $\varphi$ are given in terms of four parameters $a_\n, \left|b_\n\right|, c_\n$, and $d_\n$.
Therefore, one can solve the four given equations to get $a_\n, \left|b_\n\right|, c_\n$, and $d_\n$ in terms of the masses and the angle $\varphi$.
%%%%%%%%%%%%%%%%%%%%%%%%%%%%%%%%%%
%%%%%%%%%%%%%%%%%%%%%%%%%%%%%%%%%%%
\section{The seesaw mechanism and the $S_+$ realized $\m$--$\tau$ symmetry}
%%%%%%%%%%%%%%%%%%%%%%%%%%%%%%%%%%%
We impose now the $\m-\tau$-symmetry, defined by the matrix $S=S_+$, at the Lagrangian level within a model for the Leptons sector. Then, we shall
 invoke the type-I see-saw mechanism to address the origin of the effective neutrino mass matrix, with consequences on leptogenesis. The procedure has already been done in \cite{LCN-z23u1} for other $Z_2$-symmetries.
 \subsection{The charged lepton sector}
We start with the part of the SM Lagrangian responsible for giving masses to the charged leptons:
\bea \label{L1}
\mathcal{L}_1 =
Y_{ij} \, \overline{L} _i
\,
\phi \, \ell _j ^c,
\eea
%%%%%%%%%%%%%%%%%%%%%%%%%%%%%%%%%%%
where the SM Higgs field
$\phi$ and the right handed (RH) leptons   $\ell _j ^c$ are assumed to be singlet under $S$, whereas the left handed (LH) leptons transform in the fundamental representation of
$S$:
\bea L_i &\longrightarrow& S_{ij} L_j. \eea
%%%%%%%%%%%%%%%%%%%%%%%%%%%%%%%%%%%
Invariance under $S$ implies:
\bea S^T Y &=& Y, \eea and this forces the Yukawa couplings to have the form:
%%%%%%%%%%%%%%%%%%%%%%%%%%%%%%%%
%%%%%%%%%%%%%%%%%%%%%%%%%%%%%%%%%%%
\bea
Y &=& \left(\begin{array}{ccc} 0 & 0 & 0 \\
a & b & c \\
a & b & c \end{array}\right),
\eea
which leads, when the Higgs field acquires a vev $v$, to a charged lepton squared mass matrix of the form:
\bea
M_l
M_l ^\dagger
&=&
v^2 \,
\left(\begin{array}{ccc} 0 & 0 & 0 \\
0 & 1 & 1 \\
0 & 1 & 1 \end{array}\right)
\,
\left(|a|^2 + |b|^2 + |c|^2\right).
\eea
%%%%%%%%%%%%%%%%%%%%%%%%%%%%%%%%%%
As the eigenvectors of $M_l M_l^\dagger$ are $ \left(\,0,1/\sqrt{2},1/\sqrt{2}\,\right)^{\mbox{\tiny T}}$ with eigenvalue $2 v^2 \,\left(|a|^2 + |b|^2 + |c|^2\right)$  and
$\left(\,0,1/\sqrt{2},-1/\sqrt{2}\,\right)^{\mbox{\tiny T}}$ and $\left(~\,1,0,0\,\right)^{\mbox{\tiny T}}$ with a degenerate eigenvalue $0$, then the charged lepton mass hierarchy
can not be accommodated. Moreover, the nontrivial diagonalizing matrix, illustrated by non-canonical eigenvectors, means we are no longer in the
flavor basis. To remedy this, we introduce $SM$-singlet scalar
fields $\Delta _k$ coupled to the
lepton LH doublets through the dimension-5 operator:
%%%%%%%%%%%%%%%%%%%%%%%%%%%%%%%%%%%
\bea
\mathcal{L}_2 = \displaystyle \frac{f_{ikr}}{\Lambda}
 \, \overline{L} _i
\,
\phi \, \Delta _k \,  \ell _r ^c.
\eea
%%%%%%%%%%%%%%%%%%%%%%%%%%%%%%%%%%%
This way of adding extra SM-singlets is preferred, for suppressing flavor--changing neutral currents, than to have additional Higgs fields .
Also, we assume the $\Delta_k$'s transform under $S$ as:
\bea
\Delta _i \longrightarrow S_{ij}\, \Delta _j.
\eea
Invariance under $S$ implies, \bea
S^T f_r S= f_r,  & \mbox{where }&
\left(f_r\right)_{ij}= f_{ijr},
\eea
and thus we have the following form \bea
f_r =
\left(\begin{array}{ccc} A^r & B^r & -B^r \\
E^r & C^r & D^r \\
-E^r & D^r & C^r \end{array}
\right),
\eea
%%%%%%%%%%%%%%%%%%%%%%%%%%%%%%%%%%%
when the fields $\Delta _k$ and the neutral component of the Higgs field $\phi^\circ$ take vevs
$\left(\langle\Delta _k\rangle = \delta _k,\, v=\langle\phi^\circ\rangle\right)$ we get a charged
lepton mass matrix:
%%%%%%%%%%%%%%%%%%%%%%%%%%%%%%%%%%%
\bea
\left(M_l\right) _{ir} &=&
\displaystyle \frac{v f_{ikr} }{\Lambda} \delta _k,
\eea
%%%%%%%%%%%%%%%%%%%%%%%%%%%%%%%%
if $\delta_1, \delta _2 \ll \delta _3$ then
%%%%%%%%%%%%%%%%%%%%%%%%%%%%%%%%%%
%%%%%%%%%%%%%%%%%%%%%%%%%%%%%%%%%%%
\bea
(M_l) _{ir} \simeq
\displaystyle \frac{v f_{i3r} }{\Lambda} \delta _3
&\simeq&
\left(\begin{array}{ccc} -B^1 & -B^2 & -B^3 \\
D^1 & D^2 & D^3 \\
C^1 & C^2 & C^3 \end{array}\right),
\eea
with
$f_{13j}=-B^j$, $f_{23j}=D^j$,
$f_{33j}=C^j$ for $j=1,2,3$.
%%%%%%%%%%%%%%%%%%%%%%%%%%%%%%%%%%%
In Ref. \cite{LCN-z23u1}, a charged lepton matrix of exactly the same
form was shown to represent the lepton mass matrix in the
flavor basis with the right charged lepton mass hierarchies,
assuming just the ratios of the magnitudes of the vectors
comparable to the lepton mass ratios.
%%%%%%%%%%%%%%%%%%%%%%%%%%%%%%%%

\subsection{Neutrino mass hierarchies}

%%%%%%%%%%%%%%%%%%%%%%%%%%%%%%%%%%%%%%%%%%%%%%%%%%%%%%%%%%%%%%%%%%%%%%%%%%%%%%%%%%%%%%%%%%%%%%%%%%%%%%%%%%%%%%%%%%%%%%%%%%%%
%%%%%%%%%%%%%%%%%%%%%%%%%%%%%%%%%%%%%%%%%%%%%%%%%%%%%%%%%%%%%%%%%%%%%%%%%%%%%%%%%%%%%%%%%%%%%%%%%%%%%%%%%%%%%%%%%%%%%%%%%%%%
The effective light LH neutrino mass
matrix is generated through the seesaw mechanism formula
\bea
\label{seesaw}
M_\n&=&M_\n^D M_R^{-1} M_\n^{D\mbox{\tiny T}},
\eea
where the Dirac neutrino mass matrix $M_\n^D$
comes from the Yukawa term
\bea
\label{diracnuetrino}
g_{ij}\; \overline{L}_i\; i\tau_2\, \Phi^* \n_{Rj},
\eea
upon the Higgs field acquiring a vev, whereas the
symmetric
Majorana neutrino mass matrix $M_R$ comes from a term ($C$ is the charge conjugation matrix)
\bea
\label{majorananeutrino}
{1\over 2}\, \n_{Ri}^T\, C^{-1}\, \left(M_R\right)_{ij}\, \n_{Rj}.
\eea
We assume the RH neutrino to transform under $S$ as:
\bea
\n_{Rj } \longrightarrow S_{jr} \n_{Rr},
\eea
and thus the $S$-invariance
leads to
\bea
S^T\, g\, S = g &,& S^T\, M_R\, S = M_R.
\eea
This forces the following textures:
%%%%%%%%%%%%%%%%%%%%%%%%%%%%%%%%%%
%%%%%%%%%%%%%%%%%%%%%%%%%%%%%%%%%%%
%%%%%%%%%%%%%%%%%%%%%%%%%%%%%%%%%%
\bea
M_\n^D = v \,
\left(\begin{array}{ccc}
A_D & B_D & - B_D\\
E_D & C _ D & D _D \\
-E_D & D_D & C_D \end{array}\right) &,&
M_R = \Lambda _R \,
\left(\begin{array}{ccc}
A_R & B_R & - B_R\\
B_R & C _ R & D _R \\
-B_R & D_R & C_R \end{array}\right),
\label{formM_R}
\eea
where the explicitly appearing scales $\Lambda_R$ and $v$ characterize respectively the heavy RH Majorana neutrino masses and the electro-weak scale. Later, for numerical estimates, we shall take $\Lambda_R$ and $v$ to be respectively around $10^{14}$ GeV and $175$ GeV, so the scale characterizing the effective light neutrino $\frac{v^2}{\Lambda_R}$ would be around $0.3$ eV.  Throughout the work, where no risk of confusion, these scales will not be written explicitly in the formulae in order to simplify the notations. The resulting effective matrix $M_\n$ will have the form of Eq.(\ref{massmatrix+}) with
\bea
A_\nu &=&[
(C_R ^2 - D_R ^2) \, A_D^2
- 4 \, B_R \, (C_R + D_R) \, A_D \, B _D +
2\, A_R \, ( C_R + D_R) \, B_D ^2]/ \mbox{det}M_R,\nn
\\
%%%%%%%%%%%%%%%%%%%%%%%%%%%%%%%%
B_\nu &=&
- ( C_R + D_R) \,
\{
(D_D - C_D) \, B_D \, A_R +
(D_R - C_R) \, E_D \, A_D +
[A_D \, ( C_D - D_D) + 2 \, B_D \, E_D]\, B_R
\} / \mbox{det}M_R,\nn \\
%%%%%%%%%%%%%%%%%%%%%%%%%%%%%%%%%%
%%%%%%%%%%%%%%%%%%%%%%%%%%%%%%%%%%%
C_\nu &=&\{
(A_R \, C_R - B_R ^2 ) \, D_D ^2 +
[
- 2 \, (A_R \, D_R + B_R^2) \, C_D +
2 \, B _R \, ( C_R + D_R) \, E_D
]\, D_D \nn\\ &&
  + (A_R \, C_R - B_R^2) \, C_D^2 -
2\, B_R \, ( C_R + D_R) \, E_D \, C_D +
E_D^2 \, ( C_R^2 - D_R^2)\} / \mbox{det}M_R,\nn \\
%%%%%%%%%%%%%%%%%%%%%%%%%%%%%%%%%%%
%%%%%%%%%%%%%%%%%%%%%%%%%%%%%%%%
%%%%%%%%%%%%%%%%%%%%%%%%%%%%%%%%%%
D_ \nu &=&\{
-(A_R \, D_R + B_R^2) \,
D_D^2 +
[
- 2 \, (- A_R \, C_R + B_R^2) \, C_D -
2 \, B _R \, ( C_R + D_R) \, E_D
] \, D_D \nn \\ && -
(A_R \, D_R + B_R^2) \, C_D^2 +
2\, B_R \, ( C_R + D_R) \, E_D \, C_D -
E_D^2 \, ( C_R^2 - D_R^2)\}/ \mbox{det}M_R, \nn \\
\mbox{det}M_R&=&
(C_R + D_R) \,
[A_R \, ( C_R - D_R) - 2 \, B_R ^2 ].
\eea

Concerning the mass spectrum of the light neutrinos, it can be related to that of the RH neutrinos through the following equation connecting the product of the square eigenmasses of $M_\n$
to those of $M_D$ and $M_R$:
\be
\mbox{det}\left(M_\n^*\,M_\n\right) = \mbox{det}\left(M_\n^{D\dagger} \,M_\n^D\right)^2\; \mbox{det}\left(M_R^*\,M_R\right)^{-1}.
\label{relspec}
\ee
As was the case for the effective neutrino squared mass matrix, we choose to write:
\bea \label{MDMDdag}
 M_\n^{D\dagger} \,M_\n^D =
 \left(\begin{array}{ccc}
 a_D & b_D &  -b_D \\
b_D^* &  c_D & d_D\\
-b_D^* & d_D & c_D
\end{array}\right),
&&
M_R^* \,M_R =
 \left(\begin{array}{ccc}
 a_R & b_R &  b_R \\
b_R^* &  c_R & d_R\\
-b_R^* & d_R & c_R
\end{array}\right),
\eea
with
\bea
\begin{array}{lll}
a_D & = & \left|A_D\right|^2 + 2 \left|E_D\right|^2,\\
b_D & = & A_D^*\,B_D + E_D^*\,C_D - E_D^*\,D_D,\\
c_D & = &  \left|B_D\right|^2 +  \left|C_D\right|^2 + \left|D_D\right|^2,\\
d_D & = &  -\left|B_D\right|^2 + C_D^*\,D_D + D_D^*\,C_D,
\end{array}
\begin{array}{lll}
a_R  &= & \left|A_R\right|^2 + 2 \left|B_R\right|^2,\\
b_R & = & A_R^*\,B_R + B_R^*\,C_R - B_R^*\,D_R, \\
c_R  & = &  \left|B_R\right|^2 +  \left|C_R\right|^2 + \left|D_R\right|^2, \\
d_R  & = &  -\left|B_R\right|^2 + C_R^*\,D_R + D_R^*\,C_R.
\end{array}
  \label{abdirs+}
  \eea
so that one can write concisely the mass spectrum of
 $M_\n^*\,M_\n$, $M_R^*\,M_R$ and $M_\n^{D\dagger}\,M_\n^D$ as:
 \be
\left\{\; c_{\n,R,D} + d_{\n,R,D},\; {a_{\n,R,D} + c_{\n,R,D} - d_{\n,R,D}\over 2} \pm {1\over 2} \sqrt{\left(a_{\n,R,D}+d_{\n,R,D}-c_{\n,R,D}\right)^2 + 8\, \left|b_{\n,R,D}\right|^2}
\;\right\}.
\label{eigs+}
\ee
The mass spectrum and its hierarchy type are determined by the eigenvaules presented in Eq.(\ref{eigs+}). One of the simple realizations which can be inferred from Eq.(\ref{relspec}) is to adjust the spectrum of $M_R^*\,M_R$ so that to follow the same kind of hierarchy as $M_\n^*\,M_\n$. However, this does not necessarily imply that  $M_\n^{D\dagger}\,M_\n^D$ will behave similarly. Also, this does not exhaust all possible realizations producing the desired hierarchy and what is stated is just a mere simple possibility.

%%%%%%%%%%%%%%%%%%%%%%%%%%%%%%%%%%%%%%%%%%%%%%%%%%%%%%%%%%%%%%%%%%%%%%%%%%%%%%%%%%%%%%%%%%%%%%%%%
%%%%%%%%%%%%%%%%%%%%%%%%%%%%%%%%%%%%%%%%%%%%%%%%%%%%%%%%%%%%%%%%%%%%%%%%%%%%%%%%%%%%%%%%%%%%%%%%%
\subsection{Leptogenesis}

In this kind of models, the unitary matrix diagonalizing
$M_R$ is not necessarily diagonalizing $M_\n^D$. In fact, the Majorana and Dirac neutrino mass matrices
have different forms dictated by the $S$-symmetry and the angle $\varphi$ in Eq.(\ref{varphi}) depends on the corresponding mass parameters.
This point is critical in generating lepton asymmetry, in contrast to other symmetries \cite{LCN-z23u1} where no freedom was left for the mixing angles
leading to the same form on $M_R$ and $M_\n^D$ with identical diagonalizing matrices.
This is important when computing the CP asymmetry induced by the lightest RH neutrinos, say $N_1$, since it involves explicitly the
unitary matrix diagonalizing $M_R$:
\bea \label{epsilonleptogenesis}
\varepsilon_1 &=&
 \frac{1}{8 \, \pi  \, v^2} \,
 \frac{1}
{\left(\tilde{M}_\n^{D\dagger}  \, \tilde{M}_\n^{D}\right) _{11}} \,
\sum
_{j=2,3}
\mbox{Im} \left\{\left[\left(\tilde{M}_\n^{D\dagger}  \, \tilde{M}_\n^{D}\right) _{1j}\right]^2\right\} \,
F \left(\frac{m^2_{Rj}}{m^2_{R1}}\right).
\eea
 where   $F(x) $ is  the function   containing the one loop
vertex and self-energy corrections \cite{Loop}, and which, for  a hierarchical heavy neutrinos mass spectrum far from almost degenerate,  is given by
\begin{eqnarray}
F (x) =  \sqrt{x} \left[\frac{1}{1 - x} + 1
  - \left(1 +x \right) \ln{\left(1 +
    \frac{1}{x} \right)}\right]
\end{eqnarray}
Assuming   that there is   a strong hierarchy among RH neutrino masses with $m_{R1} << m_{R2} << m_{R3}$, the CP asymmetry can be approximated as
\begin{eqnarray}
\varepsilon_1  \simeq - 6\times 10^{-2} \,\frac{\mbox{Im} \left\{\left[\left(\tilde{M}_\n^{D\dagger}  \, \tilde{M}_\n^{D}\right) _{12}\right]^2\right\}}{v^2 \left(\tilde{M}_\n^{D\dagger}  \, \tilde{M}_\n^{D}\right) _{11}} \,\frac{m_{R1}}{m_{R2}}.
\end{eqnarray}
The matrix $\tilde{M}_\n^D$ is the Dirac neutrino mass matrix in the basis where the RH neutrinos are mass eigenstates:
\bea
\tilde{M}_\n^D =
M_\n^D \,V_R\, F_0
\eea
%%%%%%%%%%%%%%%%%%%%%%%%%%%%%%%%%%
Here $V_R$ is the unitary matrix, defined up to a phase diagonal matrix, that
diagonalizes the symmetric matrix $M_R$, and $F_0$ is a phase diagonal matrix chosen such that
the eigenvalues of $M_R$ are real and positive.

The generated baryon asymmetry  can be written as
\begin{eqnarray}
Y_B := \frac{n_B - n_{{\bar B}}}{s} \simeq 1.3 \times 10^{-3} \times \varepsilon_1\times {\cal W} ({\tilde m}, m_{R1}),~\;\;\;\;\;\; {\tilde m} = \frac{ \left(\tilde{M}_\n^{D\dagger}  \, \tilde{M}_\n^{D}\right) _{11}}{m_{R1}}
\end{eqnarray}
where $n_B, n_{{\bar B}}$ and $s$ are  the number densities of  baryons, anti-baryons, and entropy, respectively, and    ${\cal W}$ is  a  dilution  factor which accounts for the wash-out of the total lepton asymmetry due to the $\Delta L =1$ inverse decays and the lepton violating 2-2 scattering processes, and its value can  be determined  by solving the Boltzmann equation. However, analytical expressions for ${\cal W}$ have been obtained for the cases where  (${\tilde m} > 1\;eV $) and ($1\;eV> {\tilde m} > 10^{-3}\;eV$), known as the strong and the  weak  wash-out regimes respectively\cite{BPY-BD}. For instance, in the strong wash out regime (SW), ${\cal W}$ is approximated as
\bea
{\cal W}^{(SW)} \simeq \left(\frac{ 10^{-3}\; eV}{2\tilde{m}}\right)^{1.2}
\eea

 In our case where the $S$-symmetry imposes a particular form on the symmetric $M_R$ (Eq. \ref{formM_R}),
 we can take $V_R$ as being the rotation matrix $U_+$ of Eq.(\ref{U+}) corresponding to
 \be
 \t_{R\,23}= \pi /4,\; \t_{R\,12}=\varphi_R={1\over 2}
\tan^{-1}\left(\frac{2 \, \sqrt{2} \, \left|b_R\right| }
{c_R - a_R - d_R }\right),\; \t_{R\,13}=0,\; \xi_R = \mbox{Arg}\left(b_R\right).
 \ee
As to the diagonal phase matrix, $F_0= \mbox{Diag}\left(e^{-i\a_1},\,e^{-i\a_2},\,e^{-i\a_3}\right)$, it can be chosen according to Eq.(\ref{Mdiags+}) to be
\bea
\a_1 &=&\frac{1}{2}\mbox{Arg}\left[A_R\,c_{\varphi_R}^{2} - \sqrt{2}\,s_{2\varphi_R}\,e^{-i\,\xi_R}\,B_R + \left(C_R - D_R\right)\,s_{\varphi_R}^2\,e^{-2\,i\,\xi_R}\right],\nn\\
\a_2 &=& \frac{1}{2}\mbox{Arg}\left[A_R\,s_{\varphi_R}^2 + \sqrt{2}\,s_{2{\varphi_R}}\,e^{-i\,\xi_R}\,B_R + \left(C_R - D_R\right)\,c_{\varphi_R}^2\,e^{-2\,i\,\xi_R}\right],\nn\\
\a_3 &=& \frac{1}{2}\mbox{Arg}\left( c_R + d_R\right).
\eea
We assume here that the resulting mass spectrum of $M_R$ via the diagonalizing matrix $V_R F_0$ is in increasing order, otherwise one needs to
apply a suitable permutation on the columns of the latter matrix in order to get this.
Note here that had the matrix $V_R$ diagonalized $M_\n^D$, which would have meant that $N=V_R^\dagger\, M_\n^D\, V_R$ is diagonal, then we would have reached
a diagonal $\tilde{M}_\n^{D\dagger}
\tilde{M}_\n^{D}$ equaling a product of diagonal matrices, and no leptogenesis:
\bea
\tilde{M}_\n^{D\dagger}  \, \tilde{M}_\n^{D} &=& F_0^\dagger \left(V_R^\dagger M_\n^{D\dagger} V_R\right)
\left(V_R^\dagger M_\n^D V_R\right) F_0=F_0^\dagger N^\dagger N F_0
\eea
 In contrast, we get in our case:
%%%%%%%%%%%%%%%%%%%%%%%%%%%%%%%%%%
%%%%%%%%%%%%%%%%%%%%%%%%%%%%%%%%%%%
\bea
\label{m12d}
\left(\tilde{M}_\n^{D\dagger}
\tilde{M}_\n^{D}\right)_{12}
&=& e^{i(\a_1-\a_2)}\,\left[
-\sqrt{2} \,e^{i\,\xi_R} \left(A_D\, B_D^*  + E_D \, C_D^* - E_D \, D_D^* \right) s_{\varphi_R}^2
  \right.\nn \\
&&
+ \sqrt{2} \,e^{-i\,\xi_R} \left(A_D^*\, B_D  - E_D^* \, D_D + E_D^* \, C_D \right) \nn \\
&&
\left. + s_{\varphi_R}\,c_{\varphi_R}\left(-2 \left|B_D\right|^2 - \left|C_D\right|^2 - \left|D_D\right|^2 + 2\,\left|E_D\right|^2 +
\left|A_D\right|^2 + C_D^*\,D_D + D_D^*\,C_D\right)\right]
\nn\\
%%%%%%%%%%%%%%%%%%%%%%%%%%%%%%%%%%%
\left(\tilde{M}_\n^{D\dagger}
\tilde{M}_\n^{D}\right)_{13} &=&0 \nn
\\
%%%%%%%%%%%%%%%%%%%%%%%%%%%%%%%%
\left(\tilde{M}_\n^{D\dagger}
\tilde{M}_\n^{D}\right)_{11}
&=&
 c_{\varphi_R}^2\,\left( \left|A_D\right|^2 +  2 \,\left|E_D\right|^2 \right)  \nn \\
&&
+ s_{\varphi_R}^2\,\left( 2\,\left|B_D\right|^2 +  \left|C_D\right|^2 + \left|D_D\right|^2 - C_D^*\,D_D - C_D\,D_D^* \right)
\nn \\
&&
- \sqrt{2}\,s_{\varphi_R}\,c_{\varphi_R}\left( A_D\,B_D^* \, e^{i\,\xi_R} -E_D\,D_D^* \, e^{i\,\xi_R} + E_D\,C_D^* \, e^{i\,\xi_R} + \mbox{h.c}\right).
\eea
%%%%%%%%%%%%%%%%%%%%%%%%%%%%%%%%
We see that
$\left(\tilde{M}_\n^{D\dagger}\tilde{M}^{D}\right)_{12}$ is complex in general, and the question is asked whether or not one can tune it to produce the correct
CP asymmetry.
Clearly, the phase of $\left(\tilde{M}_\n^{D\dagger}\tilde{M}_\n^{D}\right)_{12}$ would be the triggering factor in producing the baryon asymmetry.
More explicitly,
\be
\mbox{Im}[\left(M^{\dagger D}_\n M^D_\n\right)_{12}]^2 \propto  \sin{\left[2\left(\phi + \alpha _1 - \alpha _2\right)\right]}
\label{eps},
\ee
where $\phi$ is the phase of the entry $\left(V_R^\dagger\, M_\n^D\, V_R\right)_{12}$.

Considering  that $m_{R1} < 10^{14}\;GeV$ and the Yukawa neutrino couplings to be not too small compared to the one which makes the see-saw mechanism more natural, which corresponds to   ${\tilde m} > 10^{-3}\;eV$, and hence the  baryon asymmetry can be expressed as

\begin{eqnarray}\label{YB}
Y_B \simeq  1.1 \times 10^{-9}\left(\frac{r_{12}}{0.1}\right) \left(\frac{m_{R1}}{10^{13}\;GeV}\right) \left( \frac{10^{-3} \mbox{eV}}{\tilde{m}}\right)^{0.2}\left[ \frac{|(M^{D \dagger}_\n M^D_\n)_{12}|}{ (M^{D \dagger}_\n M^D_\n)_{11}}\right]^2 \sin{\left[2\left(\phi+\alpha _1 - \alpha _2\right)\right]}
\label{YB}
\end{eqnarray}
with $r_{12} = m_{R1}/{m_{R2}}$ which parametrizes how strong is the hierarchy of the  RH neutrinos mass spectrum.  If the matrix elements $\left(M^{D \dagger}_\n M^D_\n\right)_{11}$ and $\left(M^{D \dagger}_\n M^D_\n\right)_{12}$  are of the same order, then, for ${\tilde m}$ of the order of  $\frac{v^2}{\Lambda_R}\simeq 0.3\;eV$, we have
\begin{eqnarray}
Y_B \simeq  0.35 \times 10^{-9}\left(\frac{r_{12}}{0.1}\right) \left(\frac{m_{R1}}{10^{13}\;GeV}\right) \sin{\left[2\left(\phi + \alpha _1 - \alpha _2\right)\right]}
\end{eqnarray}
So,  for hierarchical heavy RH neutrino mass spectrum and with $m_{R1}> 10^{13}~GeV$  one can  adjust  the value of  Majorana  phase  difference $(\alpha _1 - \alpha _2)$ to obtain $Y_B$ equals to the observed value\cite{CMB}.% = \eta^{(obs)}_B = 6 \times 10^{-10}$.

The above estimate  for  the baryon asymmetry   assumed  $|(M^{D \dagger}_\n M^D_\n)_{12}|/{(M^{D \dagger}_\n M^D_\n)_{11}} \sim 1$, and it is not generic by any mean. However, from the  equation (\ref{YB}) it is clear   that one can easily obtain a value of $Y_B$,  that is in agreement with the observation, corresponding to  many other possible choices  for the values  of   the  matrix elements   of $(M^{D \dagger}_\n M^D_\n)$, and the mass of the lightest RH neutrino \cite{LCN-z23u1}.

\section{A possible deviation from the $\m$--$\tau$ symmetry through $S_+$ and its consequences}
We saw that exact $\m$--$\tau$-symmetry implied a vanishing value for the mixing angle $\t_{13}$. Recent oscillation data pointing to a small but non-vanishing value for
this angle suggest then a deviation on the exact symmetry texture in order to account for the observed mixing. We showed in \cite{LCHN1} how ``minimal"
perturbed textures disentangling the effects of the perturbations can account for phenomenology. We shall consider now, within the scheme of type-I seesaw, a specific
perturbed texture imposed on Dirac neutrino mass matrix $M_\n^D$, and parameterized by only one small parameter $\a$, and show how it can resurface on
the effective neutrino mass matrix $M_\n$, which is known to be phenomenologically viable.
We compute then the ``perturbed" eigenmasses and mixing angles to first order in $\a$, whereas we address in the next section
 the question of finding numerically a viable pattern for $M^D_\n$ and $M^R$ leading to $M_\n$ consistent with the phenomenology.
Thus, we assume a perturbed $M_\n^D$ of the form
\bea \label{perturbedform}
M_\n^D &=&
\left(\begin{array}{ccc}
A_D & B_D \left(1+ \alpha \right) & - B_D\\
E_D & C _ D & D _D \\
-E_D & D_D & C_D \end{array}\right)
\eea
%%%%%%%%%%%%%%%%%%%%%%%%%%%%%%%%%%%
The small parameter $\a$ affects only one condition defining the exact $S$-symmetry texture, and can be expressed  as:
\bea
\a &=& -\frac{\left(M_\n^D\right)_{12}+\left(M_\n^D\right)_{13}}{\left(M_\n^D\right)_{13}}.
\eea
Applying the seesaw formula of Eq.(\ref{seesaw}) with $M_R$ given by Eq.(\ref{formM_R})  we get then:
\bea
\label{Mnu}
M_{\nu}\left(1,1\right) &=& M_{\nu}^0 \left(1,1\right) +
\alpha ^2 \frac{B_D^2 \left(C_RA_R - B_R^2\right)}
{\mbox{det}\, M_R} + \alpha \frac
{2 B_D \left(C_R+D_R\right)\, \left(A_R B_D -B_R A_D\right)}
{\mbox{det}\, M_R}
\nn\\
M_{\nu}\left(1,2\right) &= & M_{\nu}^0 \left(1,2\right) +
\alpha
\frac
{B_D \left[A_R\left(C_RC_D-D_RD_D\right)-B_R^2\left(D_D+C_D\right)-E_D B_R\left(D_R+C_R\right)\right]}
{\mbox{det}\, M_R}
\nn\\
M_{\nu} \left(1,3\right) &=&M_{\nu}^0 \left(1,3\right) +
\alpha
\frac
{ B_D \left[A_R\left(C_RD_D-D_RC_D\right)
-B_R^2\left(D_D+C_D\right)+E_DB_R\left(D_R+C_R\right)\right]}
{\mbox{det}\, M_R}
\nn\\
M_{\nu} \left(2,2\right) &=& M_{\nu}^0 \left(2,2\right) = M_{\nu}^0 \left(3,3\right) =M_{\nu} \left(3,3\right)  \nn\\
M_{\nu} \left(2,3\right) &=&M_{\nu}^0 \left(2,3\right)
\eea
where $M_{\nu}^0$ is the `unperturbed' effective neutrino mass matrix (corresponding to $\a =0$) and thus can be diagonalized by $U_+^0$ of Eq.(\ref{U+}) corresponding to the following angles,
\be
 \t_{23}^0= \pi /4,\; \t_{12}^0=\varphi^0 = {1\over 2}\tan^{-1}\left(\frac{2 \sqrt{2} \left|b_\n^0\right| }{c_\n^0 - a_\n^0 - d_\n^0}\right),\; \t_{13}^0=0,\;
 \mbox{and}\; \xi^0=\mbox{Arg}\left(b_\n^0\right),
 \ee
 Here, the superscript $0$ denotes quantities corresponding to the unperturbed effective neutrino mass matrix $M_{\nu}^0$.

The mass matrix $M_\n$ can be organized in the following form,
 \bea
 \label{pert-texture}
M_{\n} &=&
\left(\begin{array}{ccc}
A_\n & B_\n \left(1+ \chi \right) & - B_\n\\
B_\n \left(1+ \chi \right) & C_\n & D_\n \\
-B_\n & D_\n & C_\n \end{array}\right)
 \eea
where the perturbation parameter $\chi$ is given by:
\be
\label{pertparameter}
\chi = -\frac{\left(M_\n\right)_{12}+\left(M_\n\right)_{13}}{\left(M_\n\right)_{13}}.
\ee
%%%%%%%%%%%%%%%%%%%%%%%%%%%%%%%%
%%%%%%%%%%%%%%%%%%%%%%%%%%%%%%%%%%
The two parameters $\chi$ and $\alpha$ are generally complex and  of the same order provided
we do not have unnatural cancelations between the mass parameters of $M_\n^D$ and $M_R$. Nevertheless and without loss of generality,
$\a$ can be made positive and real. Furthermore, as will be explained later in our numerical investigation,
$\a$ can be adjusted to have the same value as $\left|\chi\right|$.

In order to compute the new eigenmasses and mixing angles of $M_\n$, we write it in the following form working only to first order in $\a$:
\be \label{Mnudecomp}
M_\nu =
M_\nu^0 + M_\a,
\ee
where the matrix $M_\a$ is given as,
\be \label{Malpha}
M_\a=
\left(\begin{array}{ccc} \alpha_{11} & \alpha_{12} & \alpha_{13} \\
\alpha_{12} & 0 & 0 \\
\alpha_{13} & 0 & 0 \end{array}\right),
\ee
and the non-vanishing entries of $M_\a$ are found to be,
\bea \label{alphas}
\alpha_{11}
&=&
\frac
{2 \alpha  B_D \left(C_R+D_R\right) \left(A_RB_D -B_R A_D\right)}
{\mbox{det}\, M_R},
\nn \\
\alpha _{12}
&=&
\frac
{\alpha B_D \left[A_R\left(C_RC_D-D_RD_D\right)-B_R^2\left(D_D+C_D\right)-E_DB_R\left(D_R+C_R\right)\right]}
{\mbox{det}\, M_R},
\nn \\
\alpha _{13}
&=&
\frac
{\alpha B_D \left[A_R\left(C_RD_D-D_RC_D\right)
-B_R^2\left(D_D+C_D\right)+E_DB_R\left(D_R+C_R\right)\right]}
{\mbox{det}\, M_R}.
\eea
Note here that
$M_\nu \left(1,1\right)$ gets distorted by terms
of order $\alpha$ and $\alpha ^2$. However, this will not ``perturb" the relations defining $\mu$--$\tau$ symmetry, which are expressed only through $M_\nu \left(1,2\right), M_\nu \left(1,3\right), M_\nu \left(2,2\right)$ and $M_\nu \left(3,3\right)$.

We seek now a unitary matrix $Q$ diagonalizing $M_\n^*\,M_\n$, and we write it in  the form:
\bea
Q &=& U_+^0 \left(1+ I_\epsilon\right),\;\;\; I_\epsilon =
\left(\begin{array}{ccc} 0 & \epsilon _1 & \epsilon _2 \\
-\epsilon^*_1 & 0 & \epsilon _3 \\
-\epsilon^*_2 & -\epsilon^*_3 & 0 \end{array}\right),
\eea
where $I_\epsilon$ is an antihermitian matrix due to the unitarity of $Q$.
Imposing the diagonalization condition on $M_\n^*\,M_\n$, and knowing that $U_+^0$ diagonalizes $M_\nu^{0*}\,M_\nu^0$, we have:
\bea
Q^\dagger M_\nu^*\,M_\nu Q & = & \mbox{Diag}\left(\left|M_{\n\,11}^{\mbox{\tiny Diag}}\right|^2,\left|M_{\n\,22}^{\mbox{\tiny Diag}}\right|^2, \left|M_{\n\,33}^{\mbox{\tiny Diag}}\right|^2\right),\nn\\
 U_+^{0\dagger}\, M_\nu^{0*}\,M_\nu^0\, U_+^0 & = & \mbox{Diag}\left(\left|M_{\n\,11}^{0\,\mbox{\tiny Diag}}\right|^2,\left|M_{\n\,22}^{0\,\mbox{\tiny Diag}}\right|^2, \left|M_{\n\,33}^{0\,\mbox{\tiny Diag}}\right|^2\right).
\eea
Keeping only terms up to first order in $\alpha$, which is consistent with aiming to compute $I_\epsilon$ up to this order in $\a$ and thus with dropping higher orders of $I_\epsilon$,
 we get the condition:
\bea
i,j \in \{1,2,3\}, i\neq j , \left(Q^\dagger\, M_\nu^*\,M_\nu\, Q\right)_{ij} = 0 \Longrightarrow
\left[I_\epsilon\, ,\,  M_\nu^{0\mbox{\tiny Diag}*}\,M_\nu^{0\mbox{\tiny Diag}}\right]_{ij}=
\left[U_+^{0\dagger} \left(M_\n^{0*}\,M_\alpha + M_\alpha^*\,M_\n^{0}\right)  U_+^0\right]_{ij}. &&
\eea
One  can solve analytically for $\epsilon _1$, $\epsilon _2$, $\epsilon _3$ to get:
\bea
\label{epsilons}
\epsilon _1 &=&{1\over \left|M_{\n\,22}^{0\mbox{\tiny Diag}}\right|^2 -\left|M_{\n\,11}^{0\mbox{\tiny Diag}}\right|^2 }\left\{
{1\over \sqrt{2}}\,e^{-i\,\xi^0}\,\left[ \left(\a_{13}^*-\a_{12}^*\right)\,
\left(D_\n^0 - C_\n^0\right) - A_\n^{0*}\,\left(\a_{13} - \a_{12}\right) +
2\,\a_{11}^*\,B_\n^0\right]\,c_\varphi^2 +\right.\nn\\
&& 2\,\mbox{Re}\left(\a_{11}^*\,A_\n^0\right)\,s_\varphi\,c_\varphi
-{1\over \sqrt{2}}\,e^{i\,\xi^0}\,\left[ \left(\a_{13}-\a_{12}\right)\,
\left(D_\n^{0*} - C_\n^{0*}\right) - A_\n^{0}\,\left(\a_{13}^* - \a_{12}^*\right) +
2\,\a_{11}\,B_\n^{0*}\right]\,s_\varphi^2 \left.\right\},\nn\\
%%%%%%%%%%%%%%%%%%%%%%%%%%%%%%%%%%%%%%%%%%
\epsilon _2 &=&{1\over \left|M_{\n\,33}^{0\mbox{\tiny Diag}}\right|^2 -\left|M_{\n\,11}^{0\mbox{\tiny Diag}}\right|^2 }\left\{
{1\over \sqrt{2}}\,\left[ \left(\a_{13}+\a_{12}\right)\,A_\n^{0*} +
\left(C_\n^0 + D_\n^0\right)\,\left(\a_{13}^*+\a_{12}^*\right)\right]\,c_\varphi -\right. %\nn\\
%&&
\left.
e^{i\,\xi^0}\,B_\n^{0*}\,\left(\a_{12}+\a_{13}\right)\,s_\varphi
\right\},\nn\\
%%%%%%%%%%%%%%%%%%%%%%%%%%%%%%%%%%%%%%%%%%%%%%%%%%%
\epsilon _3 &=&{1\over \left|M_{\n\,33}^{0\mbox{\tiny Diag}}\right|^2 -\left|M_{\n\,22}^{0\mbox{\tiny Diag}}\right|^2 }\left\{
{1\over \sqrt{2}}\,\left[ \left(\a_{13}+\a_{12}\right)\,A_\n^{0*} +
\left(C_\n^0 + D_\n^0\right)\,\left(\a_{13}^*+\a_{12}^*\right)\right]\,s_\varphi +\right. %\nn\\
%&&
\left.
e^{-i\,\xi^0}\,B_\n^{0*}\,\left(\a_{12}+\a_{13}\right)\,c_\varphi
\right\},\nn\\
\eea
and the resulting diagonal matrix $M_\n^{\mbox{\tiny Diag}} = Q^T M_\n Q$ is such that
\bea
M_{\n\,11}^{\mbox{\tiny Diag}} &=& M_{\n\,11}^{0\,\mbox{\tiny Diag}} + c_{\varphi^0}^2 \a_{11} - \sqrt{2}\,s_{\varphi^0}\,c_{\varphi^0}\,\left(\a_{12} - \a_{13}\right)\,e^{-i\,\xi^0},\nn\\
%%%%%%%%%%%%%%%%%%%%%%%%%%%%%%%%%%%%%%%%%%%%%%%%%%%%%%%%%%%%%%
M_{\n\,22}^{\mbox{\tiny Diag}} &=& M_{\n\,22}^{0\,\mbox{\tiny Diag}} + s_{\varphi^0}^2 \a_{11} + \sqrt{2}\,s_{\varphi^0}\,c_{\varphi^0}\,\left(\a_{12} - \a_{13}\right)\,e^{-i\,\xi^0},\nn\\
M_{\n\,33}^{\mbox{\tiny Diag}} &=& M_{\n\,33}^{0\,\mbox{\tiny Diag}}.
\eea
%%%%%%%%%%%%%%%%%%%%%%%%%%%%%%%%%%%
%%%%%%%%%%%%%%%%%%%%%%%%%%%%%%%%%%%
where the diagonalized mass matrix entries $M_{\n\,11}^{0\,\mbox{\tiny Diag}}$, $M_{\n\,22}^{0\,\mbox{\tiny Diag}}$ and $M_{\n\,33}^{0\,\mbox{\tiny Diag}}$ can be inferred from those in Eq.(\ref{Mdiags+}) to be,
\bea
M_{\n\,11}^{0\,\mbox{\tiny Diag}} &=& A_\n^0\,c_{\varphi^0}^2 - \sqrt{2}\,s_{2{\varphi^0}}\,e^{-i\,\xi^0}\,B_\n^0 + \left(C_\n^0 - D_\n^0\right)\,s_{\varphi^0}^2\,e^{-2\,i\,\xi^0},\nn\\
M_{\n\,22}^{0\,\mbox{\tiny Diag}} &=& A_\n^0\,s_{\varphi^0}^2 + \sqrt{2}\,s_{2{\varphi^0}}\,e^{-i\,\xi^0}\,B_\n^0 + \left(C_\n^0 - D_\n^0\right)\,c_{\varphi^0}^2\,e^{-2\,i\,\xi^0},\nn\\
M_{\n\,33}^{0\,\mbox{\tiny Diag}} &=& C_\n^0 + D_\n^0.
\label{diagalph}
\eea
Thus one can obtain  the squared masses up to order $\a$ as,
\bea \label{massalph}
m_1^2&=&
\left|M_{\n\,11}^{0\,\mbox{\tiny Diag}}\right|^2 -\sqrt{2}\,\mbox{Re}\left\{
e^{-i\,\xi^0}\,\left[ \left(\a_{13}^*-\a_{12}^*\right)\,
\left(D_\n^0 - C_\n^0\right) - A_\n^{0*}\,\left(\a_{13} - \a_{12}\right) +
2\,\a_{11}^*\,B_\n^0\right]\,s_\varphi\,c_\varphi\right\} + \nn\\
&& 2\,\mbox{Re}\left[ A_\n^0\,\a_{11}^*\,c_\varphi^2 + \left(\a_{12}^*-\a_{13}^*\right)\,
B_\n^0\right],\nn\\
m_2^2&=&
\left|M_{\n\,22}^{0\,\mbox{\tiny Diag}}\right|^2 +\sqrt{2}\,\mbox{Re}\left\{
e^{-i\,\xi^0}\,\left[ \left(\a_{13}^*-\a_{12}^*\right)\,
\left(D_\n^0 - C_\n^0\right) - A_\n^{0*}\,\left(\a_{13} - \a_{12}\right) +
2\,\a_{11}^*\,B_\n^0\right]\,s_\varphi\,c_\varphi\right\} + \nn\\
&& 2\,\mbox{Re}\left[ A_\n^0\,\a_{11}^*\,c_\varphi^2 + \left(\a_{12}^*-\a_{13}^*\right)\,
B_\n^0\right],\nn\\
m_3^2&=&
\left|M_{\n\,33}^{0\,\mbox{\tiny Diag}}\right|^2.
\eea

In order to extract the mixing and phase angles corresponding to $Q=U_+^0\left(1+I_\epsilon\right)$, the matrix $Q$ should be multiplied by a suitable diagonal phase matrix to
ensure that the eigenvalues of $M_\n$  are real and positive. Moreover, as mentioned before, the charged lepton fields should be properly re-phased in order that one can match the adopted parameterization in Eq.(\ref{defv}).
Thus, identifying $Q$, after having been multiplied by the diagonal phase matrix and made to have a third column of real values, with the $V_{\mbox{\tiny{PMNS}}}$ one can get
 the ``perturbed" mixing angles,
\bea \label{mixalph}
t_{12} \approx
t_{\varphi^0} \left|1+ {\epsilon_1\over t_{\varphi^0}} + \epsilon_1^*\,t_{\varphi^0} \right|,&
t_{13} \approx \left|\epsilon_2\, c_{\varphi^0}  + \epsilon_3\, s_{\varphi^0}\right|, &
t_{23} \approx
\left|1- 2\, \epsilon_2\, s_{\varphi^0}\,e^{-i\,\xi^0} + 2\, \epsilon_3 \, c_{\varphi^0}\,e^{-i\,\xi^0}\right|,
\eea
and the ``perturbed" phases
\bea \label{phaalph}
\d &\approx &2\,\pi - \xi^0 - \mbox{Arg}\left(\epsilon_1^*\,c_{\varphi^0}\,e^{-i\,\xi^0} + \epsilon_2^*\right),\nn\\
%%%%%%%%%%%%%%%%%%%%%%%%%%%%%%%%%%%%%%%
\r &\approx & \pi -\mbox{Arg}\left[\left(c_{\varphi^0} - \epsilon_1^*\,s_{\varphi^0}\right)\left(\epsilon_2^*\,c_{\varphi^0} + \epsilon_3^*\,s_{\varphi^0}\right)\right] -
{1\over 2} \mbox{Arg}\left(M_{\n\,33}^{\mbox{\tiny Diag}}\,M_{\n\,11}^{\mbox{\tiny Diag}*}\right),\nn\\
%%%%%%%%%%%%%%%%%%%%%%%%%%%%%%%%%%%%%%%%%%%%%%%%%%%
\s & \approx &\pi - \mbox{Arg}\left[\left(s_{\varphi^0} + \epsilon_1\,c_{\varphi^0}\right)\left(\epsilon_2^*\,c_{\varphi^0} + \epsilon_3^*\,s_{\varphi^0}\right)\right] -
{1\over 2} \mbox{Arg}\left(M_{\n\,33}^{\mbox{\tiny Diag}}\,M_{\n\,22}^{\mbox{\tiny Diag}*}\right).
\eea

\section{Numerical investigation for the deviation from the $S_+$-realized $\m$--$\tau$ symmetry}
The numerical investigation turns out to be quite subtle due to the huge number of involved parameters which describe the relevant mass matrices and the possible deviation.
Therefore, we start by studying numerically the perturbed mass matrix texture at the level of the effective light neutrino mass matrix, then, working backward, we
reconstruct the Dirac and Majorana neutrino mass matrices together with the parameter $\a$. For our numerical purpose, it is convenient to recast the effective neutrino light
mass matrix, by using Eqs.(\ref{diagM}-\ref{melements}), into the form,
\be
M_{\n\,ab} = \sum_{j=1}^{3}\,U_{aj}\,U_{bj}\,\la_j,
\ee
where $\la_1$, $\la_2$ and $\la_3$ are defined as,
\be
\la_1 = m_1\,e^{2 i \r} , \la_2 = m_2\,e^{2 i \s}, \la_3 = m_3.
\ee
Then the texture characterized by the deviation $\chi$, where $\chi$ is a complex parameter equal to $\left|\chi\right|\,e^{i\t}$,  can be written as
\bea
M_{\n\,12} +  M_{\n\,13}\left(1 + \chi\right) = 0 &\Rightarrow &
\sum_{j=1}^{3}
\left[U_{1j}\,U_{2j} + \left(U_{1j}\,U_{3j}\right)\left(1 + \chi\right)\right]\; \lambda_j=0,\nn\\
&\Rightarrow & A_1\,\la_1 +  A_2\,\la_2 +  A_3\,\la_3 =0, \nn\\
 M_{\n\,22} - M_{\n\,33}=0 &\Rightarrow &\sum_{j=1}^{3}
\left(U_{2j}\,U_{2j} - U_{3j}\,U_{3j}\right)\; \lambda_j=0,\nn\\
 &\Rightarrow & B_1\,\la_1 +  B_2\,\la_2 +  B_3\,\la_3 =0,
\label{texd}
\eea
where
\bea
\label{ABc}
A_j = U_{1j}\,U_{2j} + U_{1j}\,U_{3j}\left(1+\chi\right), &\mbox{and}&
B_j =  U_{2j}^2 -  U_{3j}^2, \;\;\;\;\;\;\;\; (\mbox{no sum over } j).
\eea
Then the coefficients $A$ and $B$ can be written explicitly in terms of mixing
angles and Dirac phase as,
\bea
  A_1 &=& - c_{\t_{12}}\,c_{\t_{13}} \left( c_{\t_{12}} c_{\t_{23}} s_{\t_{13}} - s_{\t_{12}} s_{\t_{23}} e^{-i\,\d}\right)\,\left(1 +
\chi\right)-c_{\t_{12}} c_{\t_{13}} \left(c_{\t_{12}} s_{\t_{23}} s_{\t_{13}} + s_{\t_{12}} c_{\t_{23}} e^{-i\,\d}\right) , \nn\\
 A_2 & = &- s_{\t_{12}}\,c_{\t_{13}} \left(s_{\t_{12}} c_{\t_{23}} s_{\t_{13}} + c_{\t_{12}} s_{\t_{23}} e^{-i\,\d}\right) \, \left(1 +
\chi\right) - s_{\t_{12}} c_{\t_{13}} \left( s_{\t_{12}} s_{\t_{23}} s_{\t_{13}} - c_{\t_{12}} c_{\t_{23}} e^{-i\,\d}\right),\nn\\
 A_3 &=& s_{\t_{13}} c_{\t_{23}} c_{\t_{13}}\,\left(1 + \chi \right) + s_{\t_{13}} s_{\t_{23}} c_{\t_{13}},\nn\\
 B_1 & =&\left(- c_{\t_{12}} c_{\t_{23}} s_{\t_{13}}
+ s_{\t_{12}} s_{\t_{23}} e^{-i\,\d}\right)^2 - \left(c_{\t_{12}} s_{\t_{23}} s_{\t_{13}} + s_{\t_{12}} c_{\t_{23}} e^{-i\,\d}\right)^2, \nn \\
 B_2 & =&  \left(s_{\t_{12}} c_{\t_{23}} s_{\t_{13}} + c_{\t_{12}} s_{\t_{23}} e^{-i\,\d}\right)^2 - \left(s_{\t_{12}} s_{\t_{23}} s_{\t_{13}} - c_{\t_{12}} c_{\t_{23}} e^{-i\,\d}\right)^2, \nn \\
 B_3 &=& c_{\t_{23}}^2 c_{\t_{13}}^2 - s_{\t_{23}}^2 c_{\t_{13}}^2.
 \label{abtexd}
 \eea
Assuming $\la_3 \neq 0$, Eqs.(\ref{texd}) can be solved to yield $\la$'s ratios as,
\bea
\frac{\la_1}{\la_3} &=&
\frac{A_3\; B_2-A_2\; B_3}
{A_2\; B_1-A_1\; B_2}, \nn \\
\frac{\la_2}{\la_3} &=& \frac{A_1\; B_3-A_3\; B_1}
{A_2\; B_1-A_1\; B_2},
\label{lam12}
\eea

From the $\la$'s ratios, one can get exact results for the mass ratios $m_{13}\equiv {m_1\over m_3}$ and $m_{23} \equiv {m_2\over m_3}$ as well as for the phases $\r$ and $\s$  in terms of
the mixing angles, remaining Dirac phase $\d$ and the parameter $\chi$. In addition, one can compute the expressions for many phenomenologically relevant quantities such as:
\bea
R_\nu \equiv   \frac{\delta m^2} {\left|\Delta
m^2\right|}, && \Sigma = \sum_{i=1}^{3} m_i.\nn\\
\langle
m\rangle_e =  \sqrt{\sum_{i=1}^{3} \displaystyle \left (
|V_{e i}|^2 m^2_i \right )}, &&
\langle m \rangle_{ee}  =  \left | m_1
V^2_{e1} + m_2 V^2_{e2} + m_3 V^2_{e3} \right | \; = \; \left | M_{\n 11} \right
|.
\label{relq}
\eea
Here, $R_\nu$ characterizes the hierarchy of the solar and atmospheric mass square differences, while the effective electron-neutrino mass $\langle
m\rangle_e $ and  the effective Majorana mass term
$\langle m \rangle_{ee} $ are sensitive to the absolute neutrino mass scales and can be respectively constrained from reactor
nuclear experiments on beta-decay kinematics and neutrinoless double-beta decay. As to the mass `sum'
parameter $\Sigma$, its upper bound can be constrained from cosmological observations. As
regards the values of  the non oscillation parameters $\me$, $\mee$ and $\Sigma$, we adopt the
less conservative 2-$\sigma$ range, as reported in \cite{fog3}
for $\me$ and $\Sigma$, and in \cite{Cuoricino} for $\mee$.
\bea
\langle m\rangle_e &<& 1.8\; \mbox{eV}, \nonumber \\
\Sigma &<& 1.19 \;\mbox{eV}, \nonumber \\
\langle m\rangle_{ee} & < & 0.34-0.78\; \mbox{eV}.
\label{nosdata}
\eea

The exact expressions turn out to be cumbersome to be presented, but for the sake of illustration, we state the relevant  expressions up to leading order in $s_{\t_{13}}$ as
\bea
 m_{13} \approx  1 + \frac{2 \, s_\d s_\t
\left|\chi\right| s_{\t_{13}}}{t_{\t_{12}} T}, && m_{23} \approx  1 - \frac{2\, t_{\t_{12}} s_\d s_\t
\left|\chi\right| s_{\t_{13}}}{T},\nn \\
\r  \approx \d + \frac{s_\d \, s_{\t_{13}}\,\left(s_{\t_{23}} c_{\t_{23}} \left|\chi\right|^2 + \left|\chi\right|\,c_\t\,\left(-c_{2\t_{23}} + s_{2\t_{23}}\right) - c_{2\t_{23}}\right)}{t_{\t_{12}}\,T}, && R_\n \approx - \frac{8\, s_\d\, \,s_\t \left|\chi\right|
\,s_{\t_{13}}}{s_{2\t_{12}}\, T}
,\\
 \s  \approx \d - \frac{s_\d \, t_{\t_{12}}\, s_{\t_{13}}\,\left(s_{\t_{23}} c_{\t_{23}} \left|\chi\right|^2 + \left|\chi\right|\,c_\t\,\left(-c_{2\t_{23}} + s_{2\t_{23}}\right) - c_{2\t_{23}}\right)}{T}, &&
 m_{23}^2 - m_{13}^2 \approx  - \frac{8\, s_\d\, \,s_\t \left|\chi\right|
\,s_{\t_{13}}}{s_{2\t_{12}}\, T}, \nn\\
\me \approx m_3 \left[ 1 +  \frac{4\, s_\t \, s_\d\, \left|\chi\right|
\,s_{\t_{13}}}{t_{2\t_{12}}\,T}\right], &&
 \mee \approx m_3 \left[ 1 +  \frac{4\, s_\t \, s_\d\, \left|\chi\right|
\,s_{\t_{13}}}{t_{2\t_{12}}\,T_1}\right].\nn
\label{mrtexd}
\eea
where $T$ is defined as,
\be
T=\left|\chi\right|^2\, s_{\t_{23}}^2 + 2\,\left|\chi\right|\,c_\t\,s_{\t_{23}}\,\left(s_{\t_{23}} - c_{\t_{23}}\right) + 1 - s_{2\t_{23}}. \label{T}
\ee
Our expansion in terms of $s_{\t_{13}}$ is justified since $s_{\t_{13}}$ is typically small for phenomenological acceptable values where the best fit for $s_{\t_{13}}\approx 0.15$. This kind of expansion in terms of $s_{\t_{13}}$, in the case of partial $\mu$--$\tau$ symmetry, has many subtle properties which were fully discussed in \cite{LCHN1} and no need to repeat them here.

For the numerical generation of $M_\n$ consistent with those relations in Eq.(\ref{texd}), we vary
$\t_{12}$, $\t_{13}$ and $\d m^2$ within their allowed ranges at the $3$--$\s$ level precision reported in Table~(\ref{fits}),
while $\t_{23}$ is varied in the range $\left[43^0,47^0\right]$ in order to keep it not far away from the value predicted upon imposing exact $\mu$--$\tau$ symmetry. The Dirac phase $\d$ and the phase $\t$ are varied in their full ranges, while the parameter $\left|\chi\right|$ characterizing the small deviation from the exact $\mu$--$\tau$ symmetry
is consistently kept  small satisfying $\left|\chi\right| \le 0.3$. Scanning randomly the 7-dim free parameter space (reading  ``random" values of $\t_{12}, \t_{23}, \t_{13}, \d,  \d m^2 , \t , \left|\chi\right|$ in their prescribed ranges), then determining the $A,B$'s coefficients (Eq. \ref{abtexd}) and producing the mass ratios and Majorana phases as determined by Eqs.(\ref{lam12}) allow us, after computing the quantities of Eq.(\ref{relq}), to confront the theoretical predictions of the texture versus the experimental constraints in Table (\ref{fits}), and whence to
figure out the admissible
7-dim parameter space region. Knowing the masses and the angles in the admissible region allows us to reconstruct the whole neutrino mass matrix $M_\n$
which, as should be stressed, is based on numerical calculations using the exact formulas in Eqs.(\ref{lam12}--\ref{relq}).

The resulting mass patterns are found to be classifiable into
three categories:
\begin{itemize}
\item Normal hierarchy: characterized by $m_1 < m_2 < m_3$ and
is denoted by ${\bf N}$ satisfying numerically the bound:
\be
\frac{m_1}{m_3} < \frac{m_2}{m_3} < 0.7
\label{nor}
\ee
\item Inverted hierarchy: characterized
by $m_3 < m_1 < m_2$ and is denoted by ${\bf I}$ satisfying the bound:
\be
\frac{m_2}{m_3} > \frac{m_1}{m_3} > 1.3
\label{inv}
\ee
\item Degenerate hierarchy (meaning quasi- degeneracy): characterized
by $m_1\approx  m_2 \approx m_3$ and is denoted by ${\bf D}$. The corresponding
numeric bound is
taken to be:
\be
0.7 < \frac{m_1}{m_3} < \frac{m_2}{m_3} < 1.3
\label{deg}
\ee
\end{itemize}
Moreover, we studied for each pattern the possibility of having a singular
(non-invertible) mass matrix
characterized by one of the masses ($m_1,
\mbox{and} \; m_3$) being equal to zero (the data prohibits
the simultaneous vanishing of two masses and thus $m_2$ can not vanish). It turns out that the violation of exact
$\mu$--$\tau$ symmetry does not allow for the singular neutrino mass matrix. The reason behind this is rather simple and
can be clarified through examining the mass ratio expressions  ${m_2\over m_3}$ and ${m_2\over m_1}$ which  respectively characterize
the cases $m_1=0$ and $m_3=0$. The mass ratio expressions can be evaluated in terms of $A$'s or $B$'s coefficients defined in Eq.(\ref{abtexd}) and can also be related to $R_\n$ leading
to the following results, for the  case $m_1=0$:
\bea
{m_2\over m_3} & = &
\begin{array}{lll}
 \left\{
\begin{array}{lll}
  \left|{A_3\over A_2}\right|&\approx& \sqrt{{\left|\chi\right|^2\, c_{\t_{23}}^2 + 2\,\left|\chi\right|\,c_\t\,c_{\t_{23}}\left(s_{\t_{23}} + c_{\t_{23}}\right) + 1 + s_{2\t_{23}}
\over \left|\chi\right|^2\, s_{\t_{23}}^2 + 2\,\left|\chi\right|\,c_\t\,c_{\t_{23}}\left(s_{\t_{23}} - c_{\t_{23}}\right) + 1 - s_{2\t_{23}}}}\,{s_{\t_{13}}\over s_{\t_{12}}\,c_{\t_{12}}} + O(s_{\t_{13}}^2),\\\\
&\approx & \sqrt{ {1 + s_{2\t_{23}}
\over  1 - s_{2\t_{23}}}}\,{s_{\t_{13}}\over s_{\t_{12}}\,c_{\t_{12}}} + O(s_{\t_{13}} \left|\chi\right|),\\\\
 \left|{B_3\over B_2}\right| & \approx & {1\over c_{\t_{12}}^2} \left( 1 + 2\,t_{\t_{12}}\,t_{2\t_{23}}\,c_\d\, s_{\t_{13}}\right) + O(s_{\t_{13}}^2),
 \end{array}
\right\} \approx \sqrt{R_\n},
\end{array}
\label{mr23}
\eea
and for the case $m_3=0$:
\bea
{m_2\over m_1} & =
\begin{array}{lll}
\left\{
\begin{array}{lll}
 \left|{A_1\over A_2}\right| &\approx & 1- {\left|\chi\right|^2\, s_{\t_{23}}\,c_{\t_{23}}\,c_\d +
\left|\chi\right|\,\left[c_\d\,c_\t\left(s_{2\t_{23}} - c_{2\t_{23}}\right) + s_\t\,s_\d\right] - c_\d\,c_{2\t_{23}}
\over \left|\chi\right|^2\, s_{\t_{23}}^2 + 2\,\left|\chi\right|\,c_\t\,s_{\t_{23}}\left(s_{\t_{23}} - c_{\t_{23}}\right) + 1 - s_{2\t_{23}}}\,{s_{\t_{13}}\over s_{\t_{12}}\,c_{\t_{12}}} + O(s_{\t_{13}}^2),\\\\
 &\approx & 1+ {c_\d\,c_{2\t_{23}}\,s_{\t_{13}}\over s_{\t_{12}}\,c_{\t_{12}}\,\left(1 - s_{2\t_{23}}\right)} + O(s_{\t_{13}} \left|\chi\right|),\\\\
 \left|{B_1\over B_2}\right| & \approx & t_{\t_{12}}^2\,\left(1 + {2\,t_{2\t_{23}}\,c_\d\,s_{\t_{13}} \over s_{\t_{12}} c_{\t_{12}}}\right) + O(s_{\t_{13}}^2),
 \end{array}
\right\}  \approx  1 + \frac{R_\n}{2}.&
\end{array}
\label{mr12}
\eea
 The mass ratio ${m_2\over m_3}$ for the case $m_1=0$  should be approximately equal to $\sqrt{R_\n}$ , which means that it should be much less than one. The expression obtained from the $A$'s, although it starts from $O(s_{\t_{13}})$, can not be  tuned to a small value compatible with $\sqrt{R_\n}$ for any admissible values for the mixing angles. The mixing
 angle $\t_{13}$ plays the decisive role in this failure for not being small enough as Table~(\ref{fits}) shows. Thus no need to examine the second expression derived from the $B$'s,
 and we conclude the impossibility of having $m_1=0$ with an approximate $\mu-\tau$-symmetry. Regarding the case $m_3=0$, the mass ratio ${m_2\over m_1}$   should be approximately equal to $\left(1 + \frac{R_\n}{2}\right)$ and accordingly would be slightly greater than one.
 Each one of the two available expressions providing the mass ratio can be separately tuned to fit the desired value within the admissible ranges for the
 mixing angles and the Dirac phase $\d$. However, the compatibility of the two expressions purports the condition, ${c_{2\t_{23}}\over 2 s_{2\t_{23}}\left(1-s_{2\t_{23}}\right)}\approx R_\n$, which
 can not be met for any admissible choice for $\t_{23}$. Our numerical study confirms this conclusion where
all the phenomenologically acceptable ranges for mixing angles and Dirac phase are scanned, but no solutions could  be found satisfying the mass constraint  expressed in
Eqs.~(\ref{mr23}--\ref{mr12}). Obviously, our conclusions remain the same when we consider the exact $\mu$--$\tau$ symmetry corresponding to $\chi=0$.
\clearpage
\begin{figure}[hbtp]
\centering
\epsfxsize=15cm
\epsfbox{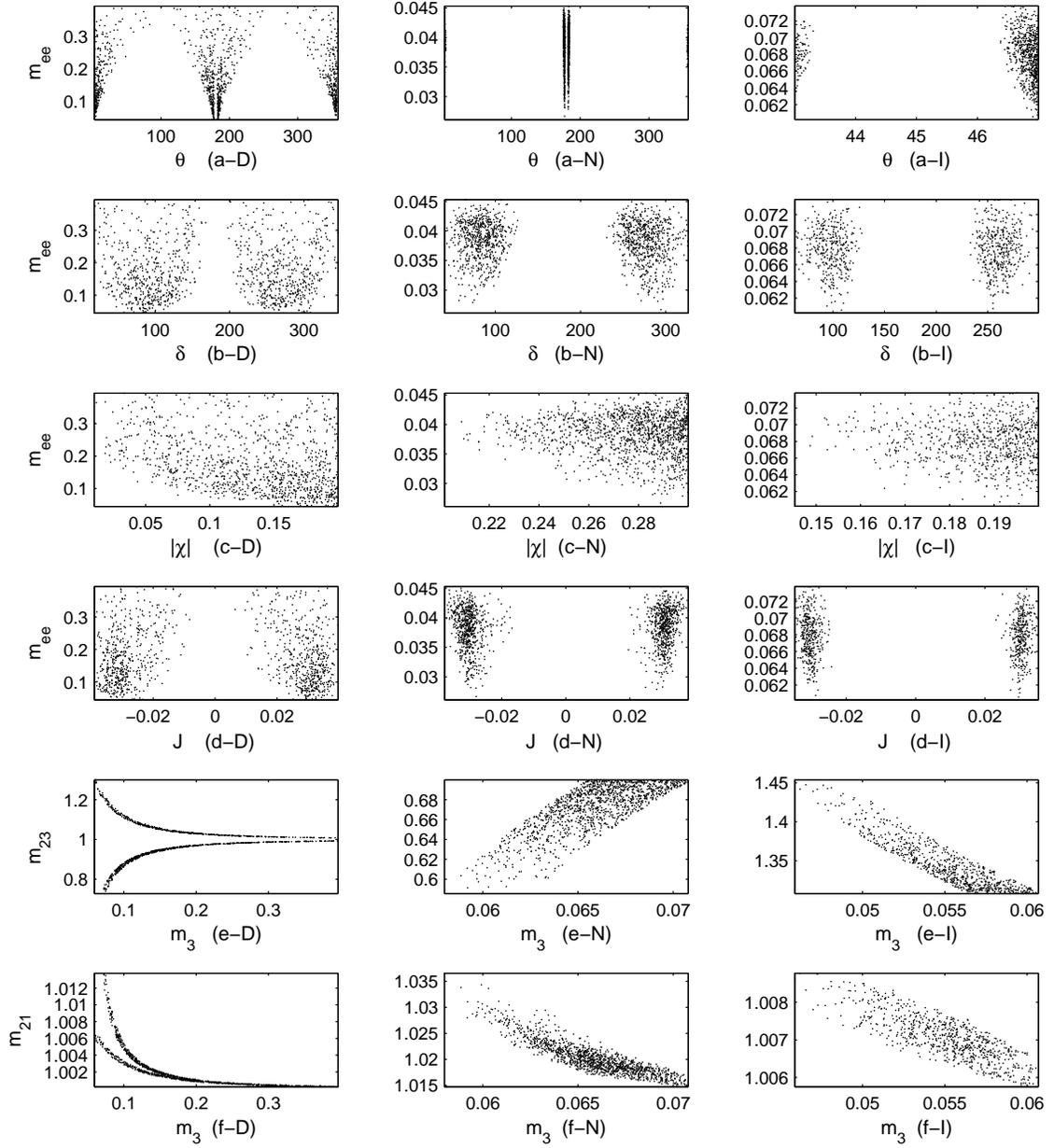}
\caption{{\footnotesize  The correlations of $\mee$ against $\t$, $\d$, $\left|\chi\right|$ and $J$ are depicted in the first four rows,
wheras the last two rows are reserved for the correlations of mass ratios $m_{23}$ and
$m_{21}$ against $m_3$.}}
\label{figone}
\end{figure}

\begin{landscape}
\begin{table}[h]
 \begin{center}
\scalebox{0.9}{
{\tiny
 \begin{tabular}{c c c c c c c c c c c c c c}
%%%%%%%%%%%%%%%%%%%%%%%%%%%%%%%%%%%%%%%%%%%%%%%%%%%%%%%%%%%%%%%%%%%%%%%%%%%%%%%
%%%%%%%%%%%%%%%%%%%%%%%%%%%%%%%%%%%%%%%%%%%%%%%Model M13 (1+x)+ M12, M22 - M33%%%%%%%%%
%%%%%%%%%%%%%%%%%%%%%%%%%%%%%%%%%%%%%%%%%%%%%%%%%%%%%%%%%%%%%%%%%%%%%%%%%%%%%%%
 \hline
 \hline
\multicolumn{14}{c}{\mbox{The pattern:} $M_{\n\,12} +
M_{\n\,13}\,\left(1+ \chi\right) =0,\; \mbox{and}\;\; M_{\n\,22} - M_{\n\,33}=0$} \\
\hline
 $\t_{12}$ & $\t_{23}$& $\t_{13}$ & $m_1$ & $m_2$& $m_3$ & $\r$ &
$\s$ & $\d$ & $\me$
 & $\mee$ & $J$ & $\left|\chi\right|$ & $\t$\\
 \hline
 \multicolumn{14}{c}{\mbox{Degenerate  Hierarchy}} \\
 \cline{1-14}
 $30.98 - 36.2$ &$[43, 44.9]\bigcup$ & $7.67 -9.94$ & $0.0521  - 0.3955$
& $0 0.0529 - 0.3955$ &
  $0.0590  - 0.3960$& $[0.003 - 14.12]\bigcup $ & $ [0.55 - 31]\bigcup$ & $[18.3 -168.1]
\bigcup$ &
  $0.0528 - 0.3954$ & $0.0452 - 0.3941$ & $ [-0.0390 -  -0.0082]\bigcup$ & $0.01 - 0.2 $ & $[2.19-83.3]\bigcup$ \\
         &  $[45.1-47]$ &  &  &  &   & $[166.3 - 179.89] $ & $ [ 154.4- 179.45]$ &$[197.7 - 345.61]$ &
   &  & $[0.0064 - 0.0397]$ &  & $[119-178.9]\bigcup$\\
   & & & & & & & & & & & & & $[181.5 - 248.3]\bigcup$\\
   & & & & & & & & & & & & & $[282.3 - 358.82]$\\
 %%%%%%%%%%%%%%%%%%%%%%%%%%%%%%%%%%%%%%%%%%%%%%%%%%%%%%%%%%%%%%%%%%%%%%%%%%%%%%%%%%%%%%%%%%%%%%%%%%%%%%%%%%
 \multicolumn{14}{c}{\mbox{Normal  Hierarchy}} \\
 \cline{1-14}
 $ 30.98 - 36.11$ &$[43, 43.78]\bigcup$ & $ 7.66 - 9.87$ & $ 0.0329 - 0.0487$
& $ 0.0329 -  0.0487$ &
  $0.0580 - 0.0708$& $[0.003 - 19.47]\bigcup $ & $ [11.04 - 35.59]\bigcup$ & $[40.44 -126.3]
\bigcup$ &
  $ 0.0339 - 0.0495$ & $ 0.0261 - 0.0453$ & $ [ -0.0379-  -0.0168]\bigcup$ & $0.2-  0.3 $ & $[2.8-4.6]\bigcup$ \\
         &  $[46.61-47]$ &  &  &  &   & $[160.2 - 179.72] $ & $ [ 144.9- 169.34]$ &$[230.7 - 326.22]$ &
   &  & $[0.0196 - 0.0382]$ &  & $[175.6-178.6]\bigcup$\\
   & & & & & & & & & & & & &$[181.8-185]\bigcup$\\
   & & & & & & & & & & & & & $[355.6 - 358.9]$\\
   %%%%%%%%%%%%%%%%%%%%%%%%%%%%%%%%%%%%%%%%%%%%%%%%%%%%%%%%%%%%%%%%%%%
   \multicolumn{14}{c}{\mbox{Inverted  Hierarchy}} \\
 \cline{1-14}
 $ 30.99 - 36.08$ &$[43, 43.31]\bigcup$ & $ 7.66-9.37$ & $ 0.0660 - 0.0790$
& $  0.0666 - 0.0795$ &
  $ 0.0459-0.0607$& $[3.35 - 10.76]\bigcup $ & $ [12.68 - 22.76]\bigcup$ & $[62.73 -127.9]\bigcup$ &
  $  0.0659 -0.0788$ & $ 0.0602 -0.0738$ & $ [  -0.0349-  -0.0243]\bigcup$ & $ 0.15 -0.2 $ & $[3.4-4.33]\bigcup$ \\
         &  $[46.38-47]$ &  &  &  &   & $[169.8 - 176.88] $ & $ [ 157.8- 168.64]$ &$[233.2 - 299.14]$ &
   &  & $[0.0232 - 0.0354]$ &  & $[175.5-177.6]\bigcup$\\
   & & & & & & & & & & & & &$[182.4-184.7]\bigcup$\\
   & & & & & & & & & & & & & $[355.5 - 356.91]$\\
 \hline
 \hline
 \end{tabular}
}}
 \end{center}
 \caption{\small  Various predictions of allowed ranges for one pattern
  violating the exact $\mu$--$\tau$ symmetry. All
the angles (masses) are
 evaluated in degrees ($eV$).}
 \label{tab2}
 \end{table}
%\end{landscape}
%%%%%%%%%%%%%%%%%%%%%%%%%%%%%%%%%%%%%%%%%%%%%%%%%%%%%%%%%%%%%%%%%%%%%%%%%%%%%%%
%%%%%%%%%%%%%%%%%%%%%%%%%%%%%%%%%%%%%%%%%%%%%%%%%%%%%%%%%%%%%%%%%%%%%%%%%%%%%%%
%\begin{landscape}
\begin{table}[h]
 \begin{center}
\scalebox{0.7}{
{\tiny
\begin{tabular}{c c c c c c c c c c c c c c c c c c c }
%%%%%%%%%%%%%%%%%%%%%%%%%%%%%%%%%%%%%%%%%%%%%%%%%%%%%%%%%%%%%%%%%%%%%%%%%%%%%%%
%%%%%%%%%%%%%%%%%%%%%%%%%%%%%%%%%%%%%%%%%%%%%%%Model M13(1+x)+ M12, M22 - M33%%%%%%%%%
%%%%%%%%%%%%%%%%%%%%%%%%%%%%%%%%%%%%%%%%%%%%%%%%%%%%%%%%%%%%%%%%%%%%%%%%%%%%%%%
 \hline
 \hline
\multicolumn{19}{c}{\mbox{Degenerate  Hierarchy}} \\
$A_\n$ & $B_\n$ & $C_\n$ & $D_\n$ &$A_R$& $B_R$ & $C_R$ & $D_R$ & $A_D$ & $B_D$ & $C_D$ & $D_D$ & $E_D$ & $\chi$ & $\alpha$  & $\t_{12}$ & $\varphi$ & $\t_{23}$ & $\t_{13}$ \\
$0.8187 + 0.0085\,i$ & $-0.0278 - 0.0300\,i$ & $0.4165 - 0.4094\,i$ & $0.3890 + 0.4097\,i$ &$ 0.8188 + 0.0086\,i$ & $-0.0297 - 0.0313\,i$ &
$0.4165 - 0.4094\,i$ & $0.3890 + 0.4097\,i$ &  $0.8187 + 0.0086\,i$ & $-0.0337 - 0.0232\,i$ & $ 0.4165 - 0.4093\,i$ & $ 0.3890 + 0.4096\,i$ &
$ -0.0238 - 0.0380\,i$ & $ 0.1089 - 0.0243\,i$ & $ 0.1116 $ &$32.63$ & $34.33$ & $44.49$ &$9.44$   \\
%%%%%%%%%%%%%%%%%%%%%%%%%%%%%%%%%%%%%%%%%%%%%%%%%%%%%%%%%%%%%%%%%%%%%%%%%%%%%%%%%%%%%%%%%%%%%%%%%%%%
$0.8045 - 0.0260\,i$ & $ -0.0229 + 0.0331\,i$ & $0.5365 + 0.3771\,i$ & $0.2557 - 0.3780\,i$ & $0.8046 - 0.0259\,i$ & $-0.0248 + 0.0366\,i$ & $0.5365 + 0.3771\,i$ & $0.2557 - 0.3780\,i$ &
$0.8046 - 0.0259\,i$ & $-0.0185 + 0.0358\,i$&  $0.5365 + 0.3771\,i$ & $0.2557 - 0.3780\,i$ & $-0.0293 + 0.0339\,i$ & $0.1960 - 0.0257i$ & $0.1977$ &
$35.81$ & $34.53$ & $44.33$ & $9.64$\\
%%%%%%%%%%%%%%%%%%%%%%%%%%%%%%%%%%%%%%%%%%%%%%%%%%%%%%%%%%%%%%%%%%%%%%%%%%%%%%%%%%%%%%%%%%%%%%%%%%%%%%%%%%%%%%%%%%%%%%%%%%%
 $0.5440 + 0.0119\,i$ & $-0.0351 - 0.0074\,i$ & $0.0152 - 0.1167\,i$ & $0.5077 + 0.1169\,i$ & $0.5441 + 0.0118\,i$ & $-0.0376 - 0.0087\,i$ &$0.0152 - 0.1167\,i$ & $0.5077 + 0.1169\,i$ &
 $0.5440 + 0.0118\,i$ & $-0.0320 - 0.0162\,i$ & $0.0152 - 0.1166\,i$ & $0.5076 + 0.1169\,i$ & $-0.0407 + 0.0002\,i$ & $ 0.1558 + 0.0417\,i$ & $0.1613$ &
 $32.50$ & $34.60$ & $44.55$  & $8.43$ \\
 %%%%%%%%%%%%%%%%%%%%%%%%%%%%%%%%%%%%%%%%%%%%%%%%%%%%%%%%%%%%%%%%%%%%%%%%%%%%%%%%%%%%%%%%%%%%%%%%%%%%%%%%%%%%%%%%%%%%%%%%%%%%%%
$\d_\n$ & $\d_\n^0$ & $\r^{\mbox{exa.}}$ & $\r^{\mbox{per}}$ & $\s^{\mbox{exa}}$ & $\s^{\mbox{per.}}$ &$m_1^0$ & $m_2^0$& $m_3^0$ & $m_1$ &
$m_2$ & $m_3$ & $m_{R3}$
 & $m_{R2}$ & $m_{R1}$ & $m_{D1}$ & $m_{D2}$ & $m_{D3}$ &   \\
 $42.36$ & $42.76$ & $1.69$ & $2.31$ & $176.94$ & $178.04$ &  $0.2511$  &  $0.2517$ & $0.2466$ & $0.2515$ & $0.2517$ & $0.2465$ &
 $8.22$ &  $8.21$ & $8.06$ & $144.22$ & $143.23$ & $140.96$ & \\
 %%%%%%%%%%%%%%%%%%%%%%%%%%%%%%%%%%%%%%%%%%%%%%%%%%%%%%%%%%%%%%%%%%%%%%%%%%%%%%%%%%%%%%%%%%%%%%%%%%%%%%%%%%%%%%%%%%%%%%%%%%%%%%%
 $ 142.75$ & $ 142.72$ & $0.68$ & $1.23$ & $ 175.93$ & $176.79$ &  $0.2469$ & $0.2475$ & $0.2426$ & $0.2473$ & $0.2475$ & $0.2424$ &
 $8.0813$ &  $8.0735$ & $7.9223$ & $141.88$ & $140.75$ & $138.64$ \\
 %%%%%%%%%%%%%%%%%%%%%%%%%%%%%%%%%%%%%%%%%%%%%%%%%%%%%%%%%%%%%%%%%%%%%%%%%%%%%%%%%%%%%%%%%%%%%%%%%%%%%%%%
  $260.66$ &  $259.97$ &  $178.79$ & $177.89$ & $5.18$ & $3.63$ &  $0.1671$ & $0.1679$ & $0.1601$ & $0.1676$ & $0.1678$ & $0.1599$ &
  $5.48$ & $5.47$ &  $5.23$ & $96.37$ & $95.15$ & $91.50$ \\
 \hline
 %%%%%%%%%%%%%%%%%%%%%%%%%%%%%%%%%%%%%%%%%%%%%%%%%%%%%%%%%%%%%%%%%%%%%%%%%%%%%%%%%%%%%%%%%%%%%%%%%%%%%%%%%%
 \multicolumn{19}{c}{\mbox{Normal  Hierarchy}} \\
 $A_\n$ & $B_\n$ & $C_\n$ & $D_\n$ &$A_R$& $B_R$ & $C_R$ & $D_R$ & $A_D$ & $B_D$ & $C_D$ & $D_D$ & $E_D$ & $\chi$ & $\alpha$  & $\t_{12}$ & $\varphi$ & $\t_{23}$ & $\t_{13}$ \\
 $0.1287 - 0.0021\,i$ & $0.0538 + 0.0038\,i$ & $0.0485 - 0.0115\,i$ & $0.1758 + 0.0115\,i$ & $0.1297 - 0.0016\,i$ & $0.0611 + 0.0040\,i$ & $0.0485 - 0.0115\,i$ & $0.1758 + 0.0115\,i$ &  $0.1294 - 0.0016\,i$ & $0.0540 + 0.0001\,i$ & $0.0485 - 0.0115\,i$ &   $0.1758 + 0.0115\,i$ &  $0.0609 + 0.0078\,i$ & $0.2700 - 0.0192\,i$ & $0.2707$ &
 $35.75$ & $33.03$ & $46.94$ &  $7.86$ \\
%%%%%%%%%%%%%%%%%%%%%%%%%%%%%%%%%%%%%%%%%%%%%%%%%%%%%%%%%%%%%%%%%%%%%%%%%%%%
 $0.1333 - 0.0148\,i$ & $0.0480 + 0.0104\,i$ & $0.0544 - 0.0355\,i$ & $0.1689 + 0.0353\,i$ & $0.1344 - 0.0142\,i$ & $0.0553 + 0.0115\,i$ & $0.0544 - 0.0355\,i$ & $0.1689 + 0.0353\,i$ &  $0.1341 - 0.0143\,i$ & $0.0486 + 0.0070\,i$ & $0.0544 - 0.0355\,i$ &   $0.1689 + 0.0353\,i$ & $0.0546 + 0.0150\,i$ & $0.2985 - 0.0213\,i$ & $0.2992$ &
 $35.44$ & $32.62$ & $46.87$ & $8.08$ \\
 %%%%%%%%%%%%%%%%%%%%%%%%%%%%%%%%%%%%%%%%%%%%%%%%%%%%%%%%%%%%%%%%%%%%%%
 $0.1325 + 0.0127\,i$ & $0.0488 - 0.0093\,i$ & $0.0537 + 0.0318\,i$ & $0.1716 - 0.0316\,i$ & $0.1337 + 0.0122\,i$ & $0.0562 - 0.0103\,i$ & $0.0537 + 0.0318\,i$ & $0.1716 - 0.0316\,i$ & $0.1334 + 0.0122\,i$ & $0.0494 - 0.0058\,i$ & $0.0538 + 0.0317\,i$ &   $0.1715 - 0.0316\,i$ & $0.0555 - 0.0140\,i$ & $0.2978 + 0.0221\,i$ & $0.2986$ &
 $36.08$ & $33.02$ & $46.84$ & $7.93$ \\
$\d_\n$ & $\d_\n^0$ & $\r^{\mbox{exa.}}$ & $\r^{\mbox{per}}$ & $\s^{\mbox{exa}}$ & $\s^{\mbox{per.}}$ &$m_1^0$ & $m_2^0$& $m_3^0$ & $m_1$ &
$m_2$ & $m_3$ & $m_{R2}$
 & $m_{R1}$ & $m_{R3}$ & $m_{D1}$ & $m_{D2}$ & $m_{D3}$ &   \\
 $97.63$ & $99.65$ & $167.90$ & $177.15$ & $ 24.17$ & $78.78$ & $0.0457$ &  $0.0461$ &    $0.0687$ & $0.0471$ & $0.0479$ & $0.0691$ & $1.57$ & $1.55$ & $2.24$ & $27.71$ &   $25.82$ & $39.24$ & \\
 %%%%%%%%%%%%%%%%%%%%%%%%%%%%%%%%%%%%%%%%%%%%%%%%%%%%%%%%%%%%%%%%%%%%%%%%%
 $82.97$ & $84.76$ &  $166.26$ & $175.94$ & $19.23$ & $99.86$ & $0.0461$ & $0.0465$    & $0.0684$ & $0.0474$ & $0.0482$ & $0.0688$ &$1.58$ & $1.55$ & $2.23$ & $27.91$ & $26.02$   & $39.08$ & \\
%%%%%%%%%%%%%%%%%%%%%%%%%%%%%%%%%%%%%%%%%%%%%%%%%%%%%%%%%%%%%%%%%%%%%%%%%%%%%%%
 $275.52$ & $273.35$ & $13.75$ & $4.18$ & $160.58$ & $88.91$ & $0.0460$ & $0.0464$ &    $0.0690$ & $0.0473$ & $0.0481$ & $0.0694$ & $1.58$ & $1.55$ & $2.25$ & $27.86$ &   $25.94$ & $39.43$ &\\
\hline
%%%%%%%%%%%%%%%%%%%%%%%%%%%%%%%%%%%%%%%%%%%%%%%%%%%%%%%%%%%%%%%%%%%
\multicolumn{19}{c}{\mbox{Inverted  Hierarchy}} \\
$A_\n$ & $B_\n$ & $C_\n$ & $D_\n$ &$A_R$& $B_R$ & $C_R$ & $D_R$ & $A_D$ & $B_D$ & $C_D$ & $D_D$ & $E_D$ & $\chi$ & $\alpha$  & $\t_{12}$ & $\varphi$ & $\t_{23}$ & $\t_{13}$ \\
$0.2322 + 0.0012\,i$ & $-0.0613 - 0.0085\,i$ & $-0.0165 - 0.0282\,i$ & $0.2113 + 0.0283\,i$ & $0.2329 + 0.0016\,i$ & $-0.0674 - 0.0090\,i$ & $-0.0165 - 0.0282\,i$ &   $0.2113 + 0.0283\,i$ & $0.2326 + 0.0016\,i$ & $-0.0617 - 0.0046\,i$ & $-0.0164 - 0.0282\,i$ & $0.2113 + 0.0283\,i$ & $-0.0669 - 0.0131\,i$ & $0.1960 - 0.0129\,i$ &
$0.1964$ & $33.63$ & $23.15$ & $43.17$ & $8.03$ \\
%%%%%%%%%%%%%%%%%%%%%%%%%%%%%%%%%%%%%%%%%%%%%%%%%%%%%%%%%%%%%%%%%%%%%%
 $0.2158 - 0.0033\,i$ & $-0.0600 - 0.0021\,i$ & $-0.0194 - 0.0058\,i$ & $0.1987 + 0.0058\,i$ & $0.2165 - 0.0030\,i$ & $-0.0658 - 0.0019\,i$ & $-0.0194 - 0.0058\,i$ &   $0.1987 + 0.0058\,i$ & $0.2162 - 0.0029\,i$ & $-0.0600 + 0.0023\,i$ & $-0.0194 - 0.0058\,i$ &  $0.1987 + 0.0058\,i$ & $-0.0657 - 0.0064\,i$ & $0.1909 - 0.0142\,i$ &
$0.1914$ & $32.66$ & $24.02$ & $43.18$ & $7.69$ \\
%%%%%%%%%%%%%%%%%%%%%%%%%%%%%%%%%%%%%%%%%%%%%%%%%%%%%%%%%%%%%%%%%%%%
 $0.2219 - 0.0043\,i$ & $-0.0603 - 0.0002\,i$ & $-0.0200 + 0.0003\,i$ & $0.2044 - 0.0004\,i$ & $0.2226 - 0.0040\,i$ & $-0.0663 + 0.0001\,i$ & $-0.0200 + 0.0003\,i$ &   $0.2044 - 0.0004\,i$ & $0.2223 - 0.0039\,i$ & $-0.0602 + 0.0040\,i$ & $-0.0199 + 0.0003\,i$ & $0.2043 - 0.0004\,i$ & $-0.0664 - 0.0041\,i$ & $0.1990 - 0.0140\,i$ &
 $0.1995$ & $35.68$ & $24.00$ & $43.16$ & $7.93$ \\
%%%%%%%%%%%%%%%%%%%%%%%%%%%%%%%%%%%%%%%%%%%%%%%%%%%%%%%%%%%%%%%%%%%%%%%%%%%%%%%%
$\d_\n$ & $\d_\n^0$ & $\r^{\mbox{exa.}}$ & $\r^{\mbox{per}}$ & $\s^{\mbox{exa}}$ & $\s^{\mbox{per.}}$ &$m_1^0$ & $m_2^0$& $m_3^0$ & $m_1$ &
$m_2$ & $m_3$ & $m_{R3}$
 & $m_{R2}$ & $m_{R1}$ & $m_{D1}$ & $m_{D2}$ & $m_{D3}$ &   \\
 $73.18$ & $62.18$ & $7.52$ & $5.98$ & $162.91$ & $158.33$ & $0.0758$ & $0.0769$ &    $0.0597$ & $0.0773$ & $0.0778$ & $0.0593$ & $2.54$ & $2.52$ & $1.95$ & $44.75$ &   $43.14$ & $34.10$ & \\
%%%%%%%%%%%%%%%%%%%%%%%%%%%%%%%%%%%%%%%%%%%%%%%%%%%%%%%%%%%%%%%%%%%%%%%%%%%%%%%%%%
$76.62$ & $67.45$ & $6.86$ & $5.94$ & $160.97$ & $157.89$ & $0.0708$ & $0.0719$ &    $0.0549$ &  $0.0723$ &  $0.0729$ & $0.0546$ & $2.38$ & $2.35$ & $1.79$ & $41.88$ &   $40.33$ & $31.38$ & \\
%%%%%%%%%%%%%%%%%%%%%%%%%%%%%%%%%%%%%%%%%%%%%%%%%%%%%%%%%%%%%%%%%%%%%%%%%%%%%%%
$81.34$ & $69.41$ & $7.58$ & $5.69$ & $163.23$ & $158.12$ & $0.0726$ & $0.0737$ &    $0.0565$ & $0.0741$ & $0.0746$ & $0.0561$ & $2.44$ & $2.41$ & $1.84$ & $42.91$ &
$41.31$ & $32.27$ & \\
\hline
\hline
 \end{tabular}
 }}
 \end{center}
 \caption{\small Numerically generated  relevant parameters for $M_\n$, $M_R$ and $M_\n^D$. Light neutrino masses are evaluated in units of eV, Dirac neutrino masses in  units of GeV, and Majorana masses in units of $10^{13}$ Gev.  The angles are
  evaluated in degrees.}
  \label{tab3}
 \end{table}
\end{landscape}

Regarding the non-singular pattern, one can deduce some restrictions concerning mixing angles and phase just by considering the approximate expression for $R_\n$ as given in Eq.(\ref{mrtexd}). The parameter $R_\n$ must be positive, non-vanishing ($R_\n\approx 0.03$) and its valued at the $3-\s$ level is reported in Table~(\ref{fits}). This clearly requires non-vanishing values for $s_{\t_{13}}$,  $s_\d$, $s_\t$  and $\left|\chi\right|$.
The nonvanishing of $s_{\t_{13}}$  implies $\t_{13} \neq 0$ which is phenomenologically favorable,  while the nonvanishing of  $s_\d$ and $s_\t$ excludes $0$, $\pi$ and $2\pi$
for both $\d$ and $\t$. The reported allowed range for $\t$ and $\d$ in Table~(\ref{tab2}) confirms these exclusions. The nonvanishing of $\left|\chi\right|$ is naturally expected otherwise there would not be  a deviation from exact $\mu$\,--\,$\tau$
 symmetry.  These  conclusions remain valid if  one  used the exact expression
for $R_\n$ instead of the first order expression. Explicit computations of $R_\n$ using its  exact expression reveal that $\t_{23}$ cannot be exactly equal to ${\pi\over 4}$, otherwise $R_\n$
would be zero, but nevertheless $\t_{23}$ can possibly stay very close to ${\pi\over 4}$, and this again is confirmed by the reported allowed values for $\t_{23}$ in Table~(\ref{tab2}).

For the sake of illustration, we show correlations involving $\mee$ against $\t$, $\d$, $\left|\chi\right|$ and $J$ where
$J$ is the Jarlskog rephasing invariant quantity which is given by $J = s_{\t_{12}}\,c_{\t_{12}}\,s_{\t_{23}}\, c_{\t_{23}}\, s_{\t_{13}}\,c_{\t_{13}}^2 \sin{\delta}$ \cite{jarlskog}.
The quantity $\mee$ is extremely important as a measure of neutrinoless double beta decay and provides a clear signature for the true nature of neutrino. The non-vanishing value for
$\mee$, if experimentally confirmed, will definitely establish the nature of neutrino as being Majorana particle. But so far, no convincing experimental evidence of the decay exists.
Other important correlations are also displayed for those involving the mass ratios $m_{12}$ and $m_{23}$ against $m_3$ which could reveal the hierarchy strength.

In Fig.~\ref{figone}, the plots (a) and (b) clearly reveal the allowed band regions
for both $\t$ and $\d$ which are quite distinct in the case of normal and inverted hierarchy, and in addition they show also the excluded region around $0$ and $\pi$.
This behavior can be mainly attributed to the constraint imposed by the parameter $R_\n$. As to the plots (c), they do not point out any clear correlation between $\mee$ and
$\left|\chi\right|$, but remarkably one can realize that in case of inverted and normal hierarchy the parameter $\left|\chi\right|$ generally tends to be larger than
what is required to be in the quasi degenerate case. Regarding the correlation of $\mee$ against $J$ (plots (d)), it is, as expected, another manifestation of
the correlation $\mee$ against $\d$, since in our investigation the size of $J$ is only controlled by $\d$ while it is apparently insensitive to the other mixing angles.
 The values of $\mee$
can not attain the zero-limit in all types of hierarchy, which is evident from
the graphs or explicitly from the corresponding covered ranges in Table~(\ref{tab2}).
There are some characteristic  features for the possible  hierarchies as can be observed from the plots (e) and (f), and  which turn out to be crucial in deriving a simple
formula for $\mee$. First, the masses $m_1$ and $m_2$ are approximately equal, as is clear
in Fig.~\ref{figone} (plots: f); second, the hierarchy is mild in both normal and inverted cases, as is evident from Fig.~\ref{figone} (plots: e-N, e-I).
The simple  approximate formula for $\mee$, capturing the essential observed features for
all kinds of hierarchies, can be deduced, assuming $m_1 \approx m_2$, from Eq.~(\ref{relq}) to be in the form:
\be
\mee  \approx  m_1\, c_{\t_{13}}^2\, \sqrt{\left[1- s_{2\t_{12}}^2\, \sin^2\left(\r-\s\right)\right]}.
\label{meeatex2d}
\ee
 The formula clearly points out that the $\mee$ scale is of the order of the scale of $m_1$($\approx m_2$) as is confirmed  from the corresponding covered ranges stated in Table~\ref{tab2}.

The numerical generation for possible $M_R$ and $M_\n^D$ for a given numerically generated $M_\n$ proceeds through the following routine (Again, this does not exhaust all possible
  $M^D_\n, M^R$ leading to the given $M_\n$). The first step consists in assuming that $M_R$ is ``proportional" to $M_\n$ but obeying exact $\mu$--$\tau$ symmetry. Thus the entries of $M_R$ can be assumed to be:
\bea
A_R = \Lambda_R\; M_{\n\,11}/v^2 = A_\n, & B_R = \Lambda_R\; \left( M_{\n\,11} - M_{\n\,13}\right)/(2 v^2) \approx B_\n, \nn\\
C_R = \Lambda_R\; M_{\n\,22}/ v^2 = C_\n, &
D_R = \Lambda_R\; M_{\n\,23}/ v^2 = D_\n,
\eea
 As said before, we took $v$ the electroweak scale characterizing the Dirac neutrino to be $175$ GeV (around the top quark mass), whereas $\Lambda_R$ the high energy scale characterizing the heavy RH Majorana neutrino is taken to be around $10^{14}$ GeV, so the scale characterizing the effective light neutrino $v^2/\Lambda_R$ would be around $0.3$ eV in agreement with data. In the second step, we assume the equality of $\a$ and $\left|\chi\right|$. Consequently,
the system of five equations given by the seesaw formula (Eq. \ref{seesaw}) applied to the symmetric matrix $M_\n$ with ($M_{\n22}=M_{\n33}$)
can then be solved for the five unknowns residing in the Dirac mass matrix having the form described in Eq.(\ref{perturbedform}). We have solved this non-linear system of equations by iteration starting with the initial guess ($A_D=A_R, B_D=B_R, C_D=C_R$ and $E_D=B_R$).

Having all parameters $A_R,\cdots, D_R$, $A_D,\cdots, E_D$ and $\a$ enables us to numerically produce the neutrino relevant quantities.
In Table~(\ref{tab3}), we report for each possible type of hierarchy three representative points containing all the parameters describing $M_\n$, $M_R$ and $M_\n^D$. In addition, the same table also contains the values of the mixing angles,  the phase angles and the masses of the light neutrinos, computed on one hand according to the exact formulae and on the other hand according to the perturbative formulae, and the two ways of computing showed good agreement. We did the perturbative calculations starting from ($M_R, M_\n^D, \a$), deduced in turn from $M_\n$ and the corresponding $\chi$, by
computing $M_\a$ (Eqs. \ref{alphas} and \ref{Malpha}) and $M_\n^0$ (Eq. \ref{Mnudecomp}) and then deducing the $\epsilon$'s (Eq. \ref{epsilons}), followed by plugging them into the perturbative formulae for the mixing angles (Eq. \ref{mixalph}), the phases (Eq. \ref{phaalph}) and the masses (Eq. \ref{massalph}).

Furthermore, the eigen masses for $M_R$ and unperturbed $M_\n^D$ are as well reported in
Table~(\ref{tab3}). We note here that we get an almost degenerate RH neutrino mass spectrum. Actually, we get for the degenerate- and inverted-hierarchy examples a mild hierarchy
in the RH eigenmasses ($m_{R1} \leq m_{R2} \simeq m_{R3}$), and so one would expect a scenario where a considerable part of the CP asymmetry is due to the decay of the lightest RH neutrino $N_1$. In
 order to estimate the baryon asymmetry in these examples one can follow the analysis of subsection 5.3 but with caution considering that we assumed there a strong hierarchy in the RH
 neutrino eigen masses leading often to $N_1$-dominated scenario. On the other hand, we obtain for the normal-hierarchy examples a mild hierarchy where the two lightest
 RH neutrinos are the almost degenerate ones ($m_{R1} \simeq m_{R2} \leq m_{R3}$), and so we would expect a scenario where the CP asymmetry is due to the decay of, at least, both $N_1$ and $N_2$.
 Here,  one should go beyond the hierarchical limit assumed in subsection 5.3 to estimate the baryon asymmetry. In \cite{branco, deppisch}, analytical formulae for the baryon asymmetry, corresponding to the case $m_{R1} \simeq m_{R2} \ll m_{R3}$, were obtained, and in \cite{blanchet} other approximate expressions, which were shown \cite{cherif} to agree well with the former ones, were derived. Although the extrapolation from the almost-degenerate two RH neutrinos case to the case of three RH neutrinos of approximately similar masses may plausibly be smooth
   regarding the fit to the Boltzmann equations, however we did not carry out the estimation of the baryon asymmetry in Table \ref{tab3} in any of the numerical examples we had, as the precise calculations go beyond
the scope of the paper and the formulated expressions are approximate, so one needs a more
refined analysis in order to draw conclusions.   Nonetheless, we have checked our assumption that the $\epsilon$'s (Eqs. \ref{epsilons}) are far smaller than $1$ in accordance with them being as perturbative factors.
%%%%%%%%%%%%%%%%%%%%%%%%%%%%%%%%%%%%%%%%%%%%%%%%%%%%%%%%%%%%%%%%%%%%%%%%%%%%%%%%%%%%%%%
\section{Realization of perturbed texture}
As we saw, perturbed textures are needed in order to account for phenomenology. We have two ways to seek models for achieving these perturbations. The first method consists of
introducing a term in the Lagrangian which breaks explicitly the symmetry \cite{Ross-Hall}, and then of expressing the new perturbed texture in terms of this breaking term. The second method
 is to keep
assuming the exact symmetry, but then we break it spontaneously by introducing new matter and enlarging the symmetry. We follow here the second approach in order to find a realization of the forms given in
Eq.(\ref{perturbedform}) for $M_D$ and in Eq.(\ref{formM_R}) for $M_R$, while assuring that we work in the flavor basis. However,  for the sake of minimum added matter, we shall not force the most general forms of $M_R$ and $M_D$, but rather be content with special forms of them leading to  an effective mass matrix $M_\n$ of the desired perturbed texture (Eq. \ref{pert-texture}).  In \cite{LCHN1} a realization was given for a perturbed texture corresponding to the $S_-$-symmetry, whereas here we  treat the
more phenomenologically motivated $S_+$-symmetry (we shall drop henceforth the $+$suffix). We present two ways, not meant by whatsoever to be restrictive but rather should be looked at as proof of existence tools, to get the three required conditions of a ``perturbed" $M_D$, non-perturbed $M_R$ and diagonal $M_l M_l^\dagger$. Both ways add new matter, but  whereas the first approach adds just a $(Z_2)^2$ factor to the $S-$symmetry while requiring some Yukawa couplings to vanish, the second approach enlarges the symmetry larger to $S \times Z_8$ but without need to equate Yukawa couplings to zero by hand.
Some ``form invariance" relations are in order:
\bea
\label{FI1}
\left\{ \left(M=M^{\mbox{\textsc{t}}}\right)\,\wedge \left[ S^{\mbox{\textsc{t}}} \cdot M \cdot S=M \right] \right\}
& \Leftrightarrow & { \left[  M = \pmatrix{A & B & -B
\cr B & C &  D\cr
-B &  D & C}
\right]},
\eea
\bea
\label{FI2}
\left\{ \left(M=M^{\mbox{\textsc{t}}}\right)\,\wedge \left[ S^{\mbox{\textsc{t}}} \cdot M \cdot S=-M \right] \right\}
& \Leftrightarrow & { \left[  M = \pmatrix{0 & B & B
\cr B & C &  0\cr
B &  0 & -C}
\right]},
\eea
\bea
\label{FI1g}
 \left[ S^{\mbox{\textsc{t}}} \cdot M \cdot S=M \right]
& \Leftrightarrow & { \left[  M = \pmatrix{A & B & -B
\cr E & C &  D\cr
-E &  D & C}
\right]},
\eea
\bea
\label{FI2g}
 \left[ S^{\mbox{\textsc{t}}} \cdot M \cdot S=-M \right]
& \Leftrightarrow & { \left[  M = \pmatrix{0 & B & B
\cr E & C &  D \cr
E &  -D & -C}
\right]},
\eea
We denote $L^\textsc{t} =  (L_1,L_2,L_3)$ with $L_i$'s,$(i=1,2,3)$ are the components of the $i^{th}$-family LH lepton doublets (we shall adopt this notation of `vectors' in flavor space even for other fields, like $l^c$ the RH charged lepton singlets, $\n_R$ the RH neutrinos, $\ldots$).

\subsection{ $S \times Z_2 \times Z_2^\prime$-flavor symmetry}

\begin{itemize}
%%%%%%%%%%%%%%%%%%%%%%%%
%%%%%%%%%%%%%%%%%%%%%%%%
\item{\bf Matter content and symmetry transformations}

We have three SM-like Higgs doublets ($\phi_i$, $i=1,2,3$) which would give mass to the charged leptons and another three Higgs doublets ($\phi^\prime_i$, $i=1,2,3$) for the
Dirac neutrino mass matrix.  All the fields are invariant under $Z_2^\prime$ except the fields $\phi^\prime$ and $\n_R$ which are multiplied by $-1$, so that we
assure that neither $\phi$ can contribute to $M_D$, nor $\phi^\prime$ to $M_l$.  The fields transformatios are as follows.
%%%%%%%%%%%%%%
\bea  &\n_R \stackrel{Z_2}{\longrightarrow} \mbox{diag}\left(1,-1,-1\right) \n_R ,\;\; \phi^\prime \stackrel{Z_2}{\longrightarrow} \mbox{diag}\left(1,-1,-1\right)  \phi^\prime,
\label{typeIZ2one} &\\
&L \stackrel{Z_2}{\longrightarrow} \mbox{diag}\left(1,-1,-1\right)  L,\;\; l^c \stackrel{Z_2}{\longrightarrow} \mbox{diag}\left(1,1,-1\right)  l^c , \;\; \phi \stackrel{Z_2}{\longrightarrow} \mbox{diag}\left(1,-1,-1\right) \phi, \label{typeIZ2two}
 \eea
\bea  &\n_R \stackrel{S}{\longrightarrow} S \n_R , \;\; \phi^\prime \stackrel{S}{\longrightarrow} \mbox{diag}\left(1,1,-1\right)  \phi^\prime,
 \label{typeISone} &\\
&L \stackrel{S}{\longrightarrow} S L, \;\; l^c \stackrel{S}{\longrightarrow}  l^c , \;\; \phi \stackrel{S}{\longrightarrow} S\phi, \label{typeIStwo}
 \eea
 %%%%%%%%%%%%%%%%%
 %%%%%%%%%%%%%%%%%
\item{\bf Charged lepton mass matrix-flavor basis}

The Lagrangian responsible for $M_l$ is given by:
\bea
 \label{L2}
 {\cal{L}}_2 &=& f^j_{ik} \overline{L}_i \phi_k  l^c_j \,
 \eea
The transformations under $S$ and $Z_2$, with the ``form invariance" relations Eqs. (\ref{FI1}--\ref{FI2g}), lead to:
\bea
f^{(1)}  =
\left(
\begin {array}{ccc}
A^1 &0& 0\\
0& C^1& D^1\\
0&D^1&C^1
\end {array}\right) , f^{(2)}  =
\left(
\begin {array}{ccc}
A^2 &0& 0\\
0& C^2& D^2\\
0&D^2&C^2
\end {array}\right), f^{(3)}  =
\left(
\begin {array}{ccc}
0 &B^3& -B^3\\
E^3& 0& 0\\
-E^3&0&0
\end {array}\right)\eea
where $f^j_{ik}$ is the ${(i,k)}^{\mbox{th}}$-entry of the matrix $f^{(j)}$.
Assuming ($v_3 \gg v_1, v_2$) we get:
\bea
M_l  = v_3
\left(
\begin {array}{ccc}
0 &0& -B^3\\
D^1& D^2 & 0\\
C^1&C^2&0
\end {array}
\right) &\Rightarrow&
M_l\;M_l^\dagger
    = v_3^2
    \pmatrix {|{\bf B}|^2  & 0 & 0 \cr
     0 & |{\bf D}|^2 & {\bf D} \cdot {\bf C}\cr
    0 & {\bf C} \cdot {\bf D} & |{\bf C}|^2},
\label{ChargedMasstypeII3Higgslc2}
\eea
where ${\bf B}=\left(0,0,-B^3\right)^T$, ${\bf D}=\left(D^1,D^2,0\right)^T$ and ${\bf C}=\left(C^1,C^2,0\right)^T$, and where the dot product is defined as ${\bf D} \cdot {\bf C} = \sum_{i=1}^{i=3}
D^iC^{i*}$. Under the reasonable assumption that the magnitudes of the Yaukawa couplings come in ratios proportional to the lepton mass ratios as $\left|B\right| : \left|C\right| : \left|D\right|
\sim m_e : m_\mu : \m_\tau$, one can show, as was done in \cite{LCHN1}, that the diagonalization of the charged lepton mass matrix can be achieved by  infinitesimally rotating
the LH charged lepton fields, which justifies working in the flavor basis to a good approximation.

%%%%%%%%%%%%%%%%%%%%%%%%%%%%
%%%%%%%%%%%%%%%%%%%%%%%%%%%%%
\item{\bf Majorana neutrino mass matrix}\\
The mass term is directly present in the Lagrangian
\bea
 \label{L_R}
 {\cal{L}}_R &=& M_{R\,ij}\, \n_{Ri}\,  \n_{Rj} \,\, \label{csawLR}.
 \eea
The invariance under $Z_2'$ is trivially satisfied while the one under
$S\times Z_2$ is more involved. The symmetry $S$ constrains $M_R$ to satisfy
\be
S^T\, M_R\, S = M_R,
\label{rest1}
\ee
whereas the restrictions due to $Z_2$ are imprinted in
the bilinear  of $\n_{Ri}\,  \n_{Rj}$ determining their
transformations under $Z_2$ as:
\bea
 \nu_{Ri}\, \nu_{Rj} \stackrel{Z_2}{\sim} B &=& \left(
\begin {array}{ccc}
1&-1 & -1\\
-1 & 1 & 1\\
-1 &1 &1
\end {array}
\right)
\label{rest2}
\eea
which means:
\bea \nu_{Ri}\, \nu_{Rj}  \stackrel{Z_2}{\longrightarrow} Z_2(\nu_{Ri}\, \nu_{Rj} )  &=& B_{ij} \nu_{Ri}\, \nu_{Rj}  (\mbox{no sum})
\eea
Thus the symmetry through Eqs.(\ref{FI1},\ref{rest1},\ref{rest2}) entails that $M_R$ would
assume the following form,
\bea
M_R & = & \label{mrs}
\left(
\begin {array}{ccc}
A_R &0& 0\\
0& C_R& D_R\\
0&D_R &C_R
\end {array}
\right).
\eea which is of the general form (Eq. \ref{formM_R}) with $B_R=0$.
%%%%%%%%%%%%%%%%%%%%%%%%%%%%%%%
%%%%%%%%%%%%%%%%%%%%%%%%%%%%%%%

\item{\bf Dirac neutrino mass matrix}

The Lagrangian responsible for the neutrino mass matrix is
\bea
 \label{L_D}
 {\cal{L}}_D &=& g^k_{ij} \overline{L}_i \tilde{\phi^\prime}_k  \n_{Rj} \,\, ,\mbox{ where  } \tilde{\phi^\prime} = i \s_2  \phi^{\prime *}
 \eea
Because of the fields transformations under $S$ and $Z_2$ we get:
\bea S^{\mbox{\textsc{t}}} g^{(k=1,2)} S = g^{(k=1,2)} &,& S^{\mbox{\textsc{t}}} g^{(k=3)} S = - g^{(k=3)}, \overline{L}_i  \n_{Rj}
 \stackrel{Z_2}{\sim} \left(
\begin {array}{ccc}
1&-1 & -1\\
-1 & 1 & 1\\
-1 &1 &1
\end {array}
\right)
\label{typeIinvariance}
\eea
where $g^{(k)}$ is the matrix whose $(i,j)^{th}$-entry is the Yukawa coupling $g^k_{ij}$. Then, the ``form invariance" relations (Eqs.\ref{FI1}--\ref{FI2g}) lead
to:
\bea
g^{(1)}  =
\left(
\begin {array}{ccc}
{\cal A}^1 &0& 0\\
0& {\cal C}^1& {\cal D}^1\\
0&{\cal D}^1&{\cal C}^1
\end {array}\right) , g^{(2)}  =
\left(
\begin {array}{ccc}
0 &{\cal B}^2& -{\cal B}^2\\
{\cal E}^2& 0& 0\\
-{\cal E}^2&0&0
\end {array}\right), g^{(3)}  = \left(
\begin {array}{ccc}
0 &{\cal B}^3& {\cal B}^3\\
{\cal E}^3& 0& 0\\
{\cal E}^3&0&0
\end {array}\right)\eea
Upon acquiring vevs ($v_i^\prime$, $i=1,2,3$) for the Higgs fields ($\phi^\prime_i$), we get for Dirac neutrino mass matrix the form:
\bea
M_D  &=&\left(
\begin {array}{ccc}
v_1'\, {\cal A}^1 & v_2'\, {\cal B}^2 + v_3'\, {\cal B}^3 & -v_2'\, {\cal B}^2 + v_3'\, {\cal B}^3\\
v_2'\, {\cal E}^2 + v_3'\, {\cal E}^3 & v_1'\, {\cal C}^1 & v_1'\, {\cal D}^1 \\
-v_2'\, {\cal E}^2 + v_3'\, {\cal E}^3 & v_1'\, {\cal D}^1 & v_1'\, {\cal C}^1
\end{array}
\right),
\label{MD1}
\eea
which can be put into the form,
\bea
M_D & =& \left(
\begin {array}{ccc}
A_D &B_D \left(1+\a\right)& -B_D \\
E_D \left(1+\beta\right)& C_D& D_D\\
-E_D &D_D&C_D
\end {array}
\right).
\label{MD2}
\eea
with
\bea
\a = \frac{2v^\prime_3 {\cal B}^3}{v^\prime_2 {\cal B}^2 - v^\prime_3 {\cal B}^3} &,& \b = \frac{2v^\prime_3 {\cal E}^3}{v^\prime_2 {\cal E}^2 - v^\prime_3 {\cal E}^3}.
\label{alfabeta}
\eea
If the vevs satisfy $v^\prime_3 \ll v^\prime_2$ and the Yukawa couplings are of the same order, then we get perturbative  parameters $\a, \beta \ll 1$.

The deformations appearing in the Dirac mass matrix as described in Eqs.(\ref{MD1}--\ref{alfabeta}) would resurface in the effective light neutrino mass matrix $M_\nu$ through the seesaw formula (Eq.\ref{seesaw}) with $M_R$ given in Eq.(\ref{mrs}). The resulting deformations in $M_\n$ can be described by two parameters:
\bea
\chi \equiv -\frac{M_\n\left(1,2\right) +  M_\n\left(1,3\right)}{M_\n\left(1,3\right)} &,&
\xi \equiv \frac{M_\n\left(2,2\right) -  M_\n\left(3,3\right)}{M_\n\left(3,3\right)}.
\label{defdef1}
\eea
One can repeat now the analysis of the last subsection in order to compute $\chi, \xi$ in terms of $\a, \b$  and other mass parameters to get:
\bea
\chi &=& -\frac{\a\, A_R B_D \left(C_R -D_R\right) \left(C_D + D_D\right) + \beta A_D E_D \left(C_R^2 -D_R^2\right)}{\a A_R B_D \left(C_R D_D -D_R C_D\right) + B_D A_R \left(D_R + C_R\right) \left(D_D - C_D\right) - E_D A_D \left(C_R^2 -D_R^2\right)}\nn\\
\xi &=& \frac{\beta \left(\beta-2\right) E_D^2 \left(C_R^2 -D_R^2\right)}{A_R \left[ C_R \left(D_D^2 + C_D^2\right) - 2 C_D D_D D_R\right] + E_D^2 \left(C_R^2 -D_R^2\right)},
\label{defdef2}
\eea
We note here that we do not get in general the desired pattern (Eq. \ref{pert-texture}) corresponding to disentanglement of the perturbations ($\xi =0$). However, for specific choices of Yukawa couplings, for e.g. ${\cal E}^3=0$ leading to $\b=0$ and hence $\xi =0$, we get this form, in which case $M_D$ is of the form  (Eq.\ref{perturbedform}) and $\chi$ of Eq.(\ref{defdef2}) would
also be given by Eq.(\ref{pertparameter}) with $B_R=0$.
\end{itemize}

\subsection{ $S \times Z_8 $-flavor symmetry}

\begin{itemize}
%%%%%%%%%%%%%%%%%%%%%%%%%%%%%%%%%%%%%
%%%%%%%%%%%%%%%%%%%%%%%%%%%%%%%%%%%%%
\item{\bf Matter content and symmetry transformations}

 In addition to the left doublets ($L_i$, $i=1,2,3$), the RH charged singlets ($l^c_j$, $j=1,2,3$), the RH neutrinos ($\n_{Rj}$, $j=1,2,3$) and the SM-Higgs three doublets ($\phi_i$, $i=1,2,3$) responsible for the charged lepton masses, we have now four Higgs doublets ($\phi^\prime_j$, $j=1,2,3,4$) giving rise when acquiring a vev to Dirac neutrino mass matrix, and also two Higgs singlet scalars
($\Delta_k$, $k=1,2$) related to Majorana neutrino mass matrix. We denote the octic root of the unity by $\omega=e^{\frac{i\pi}{4}}$. The fields transform as follows.
\bea
&L \stackrel{S}{\longrightarrow} S L ,\;\; l^c \stackrel{S}{\longrightarrow}  l^c ,\;\;  \phi \stackrel{S}{\longrightarrow} S\phi, & \label{csawSone} \\
&\n_R \stackrel{S}{\longrightarrow} S \n_R ,\;\; \phi^\prime \stackrel{S}{\longrightarrow} \mbox{diag}\left(1,1,1,-1\right)  \phi^\prime, \;\; \Delta  \stackrel{S}{\longrightarrow} \Delta & \label{csawStwo}
\eea
\bea
&L \stackrel{Z_8}{\longrightarrow} \mbox{diag}\left(1,-1,-1\right)  L, \;\;l^c \stackrel{Z_8}{\longrightarrow} \mbox{diag}\left(1,1,-1\right)  l^c , \;\; \phi \stackrel{Z_8}{\longrightarrow} \mbox{diag}\left(1,-1,-1\right) \phi,& \label{csawZ8one}\\
 &\n_R \stackrel{Z_8}{\longrightarrow} \mbox{diag}\left(\omega,\omega^3,\omega^3\right) \n_R ,\;\; \phi^\prime \stackrel{Z_8}{\longrightarrow} \mbox{diag}\left(\omega,\omega^3,\omega^7,\omega^3\right)  \phi^\prime,\;\;
 \Delta  \stackrel{Z_8}{\longrightarrow} \mbox{diag}\left(\omega^6,\omega^2\right)\Delta
 \label{csawZ8two} &
 \eea
Note here that we have the following transformation rule for $\tilde{\phi^\prime} \equiv i \s_2 \phi^{\prime *}$:
\bea \label{tilderule}
\tilde{\phi^\prime}  \stackrel{S}{\longrightarrow} \mbox{diag}(1,1,1,-1) \tilde{\phi^\prime} &,& \tilde{\phi^\prime}  \stackrel{Z_8}{\longrightarrow} \mbox{diag}(\omega^7,\omega^5,\omega,\omega^5) \tilde{\phi^\prime}
\eea
%%%%%%%%%%%%%%%%%%%%%%%%%%%%%%%%%%%%%%%
%%%%%%%%%%%%%%%%%%%%%%%%%%%%%%%%%%%%%%%
\item{\bf Charged lepton mass matrix-flavor basis}

The symmetry restriction in constructing  the charged lepton mass Lagrangian as given by Eq.(\ref{L2}) is similar to what is obtained in the case of ($S\times Z_2 \times Z_2'$). The similarity orginates from the fact that the charges assigned to the fields  ($L,l^c,\phi$) corresponding to the factor $Z_2$ (of $S\times Z_2 \times Z_2'$ ) and that of $Z_8$ (of $S\times Z_8$) are the same. Thus we end up, assuming again a hierarchy in the Higgs $\phi$'s fields vevs ($v_3 \gg v_2,v_1$), with a charged lepton mass matrix adjustable to be approximately in the flavor basis. Note also here that the symmetry forbids the term
$\overline{L}_i \phi^\prime_k  l^c_j$ since we have:
\bea \overline{L}_i  l^c_j
 \stackrel{Z_8}{\sim} \left(
\begin {array}{ccc}
1&1 & -1\\
-1 & -1 & 1\\
-1 &-1 &1
\end {array}
\right) &\stackrel{\mbox{Eq.}\ref{csawZ8two}}{\Longrightarrow}& \nexists i,j,k: \overline{L}_i \phi^\prime_k  l^c_j=Z_8(\overline{L}_i \phi^\prime_k  l^c_j)
\label{noterm}\eea
%%%%%%%%%%%%%%%%%%%%%%%%%%%%%%%%%%%%%%
%%%%%%%%%%%%%%%%%%%%%%%%%%%%%%%%%%%%%%
\item{\bf Dirac neutrino mass matrix}

The Lagrangian responsible for the Dirac neutrino mass matrix is given by Eq. (\ref{L_D}). By means of fields transformations we have:
\bea S^{\mbox{\textsc{t}}} g^{(k=1,2,3)} S = g^{(k=1,2,3)} &,& S^{\mbox{\textsc{t}}} g^{(k=4)} S = - g^{(k=4)}, \overline{L}_i  \n_{Rj}
 \stackrel{Z_8}{\sim} \left(
\begin {array}{ccc}
\omega&\omega^3 & \omega^3\\
\omega^5 & \omega^7 & \omega^7\\
\omega^5 &\omega^7 &\omega^7
\end {array}
\right)
\label{csawinvariance}
\eea
where $g^{(k)}$ is the matrix whose $(i,j)^{th}$-entry is the Yukawa coupling $g^k_{ij}$. Then,  the ``form invariance" relations impose the following forms:
\bea &
g^{(1)}  =
\left(
\begin {array}{ccc}
{\cal A}^1 &0& 0\\
0& 0& 0\\
0&0&0
\end {array}\right) , g^{(2)}  =
\left(
\begin {array}{ccc}
0 &{\cal B}^2& -{\cal B}^2\\
0& 0& 0\\
0&0&0
\end {array}\right), g^{(3)}  = \left(
\begin {array}{ccc}
0 &0& 0\\
0& {\cal C}^3& {\cal D}^3\\
0&{\cal D}^3&{\cal C}^3
\end {array}\right),&\nn\\&
g^{(4)}  =
\left(
\begin {array}{ccc}
0 &{\cal B}^4&{\cal B}^4\\
0& 0& 0\\
0&0&0
\end {array}\right),
&
\eea
When the Higgs fields ($\phi^\prime_i$) get vevs ($v_i^\prime$, $i=1,2,3,4$), we obtain:
\bea
M_D  = \Sigma_{k=1}^{k=4} v^\prime_k g^{(k)} &=& \left(
\begin {array}{ccc}
v_1' {\cal A}^1 & v_2' {\cal B}^2 +  v_4' {\cal B}^4 & - v_2' {\cal B}^2 +  v_4' {\cal B}^4 \\
0 & v_3' {\cal C}^3 & v_3' {\cal D}^3\\
0 & v_3' {\cal D}^3 & v_3' {\cal C}^3\\
\end {array}\right), \label{csawDiracMatrix}
\eea
which is of the form of Eq.(\ref{perturbedform}) with $E_D=0$:
\bea
M_D & =& \left(
\begin {array}{ccc}
A_D &B_D \left(1+\a\right)& -B_D \\
0 & C_D& D_D\\
0 &D_D&C_D
\end {array}
\right),
\label{MD3}
\eea
where
\bea
\a &=& \frac{2v^\prime_4 {\cal B}^4}{v^\prime_2 {\cal B}^2 - v^\prime_4 {\cal B}^4}
\label{alfa}
\eea
If the vevs satisfy $v^\prime_4 \ll v^\prime_2$ and the Yukawa couplings are of the same order then we get a perturbative  parameter $\a \ll 1$.
%%%%%%%%%%%%%%%%%%%%%%%%%%%%%%%%%%%%
%%%%%%%%%%%%%%%%%%%%%%%%%%%%%%%%%%%%
\item{\bf Majorana neutrino mass matrix}

The mass term is generated from the Lagrangian
\bea
 \label{L_R}
 {\cal{L}}_R &=& h^k_{ij}\, \Delta_k\, \n_{Ri}\,  \n_{Rj} \,\, \label{csawLR}
 \eea
 Under $Z_8$ we have the bilinear:
\bea &\n_{Ri}\;  \n_{Rj}
 \stackrel{Z_8}{\sim} \left(
\begin {array}{ccc}
\omega^2&\omega^4 & \omega^4\\
\omega^4 & \omega^6 & \omega^6\\
\omega^4 &\omega^6 &\omega^6
\end {array}
\right) \stackrel{\mbox{Eq.}\ref{csawZ8two}}{\Longrightarrow}&  \nonumber \\&{\cal{L}}_R=h^1_{11}\, \Delta_1\, \n_{R1}\,  \n_{R1}
+ h^2_{22}\, \Delta_2\, \n_{R2}\,  \n_{R2} + h^2_{23}\, \Delta_2\, \n_{R2}\,  \n_{R3} + h^2_{32}\, \Delta_2\, \n_{R3}\,  \n_{R2} + h^2_{33}\, \Delta_2\, \n_{R3}\,  \n_{R3}
&\;\;\;\;
\label{csawZ8lag}
\eea
If we call $h^{(k)}$  the matrix whose $(i,j)^{th}$-entry is the coupling $h^k_{ij}$ then we have (the cross sign denote a non-vanishing entry):
\bea &
h^{(1)}  =
\left(
\begin {array}{ccc}
\times &0& 0\\
0& 0& 0\\
0&0&0
\end {array}
\right) , h^{(2)}  =
\left(
\begin {array}{ccc}
0 &0& 0\\
0& \times& \times\\
0&\times&\times
\end {array}
\right),&
\label{csawZ8mass}
\eea
Then the ``form invariance" relations lead to:
\bea & S^{\mbox{\textsc{t}}} h^{(k)} S = h^{(k)} ,
 \stackrel{\mbox{Eqs.}\ref{FI1},\ref{csawZ8mass}}{\Longrightarrow}& \nn \\ & h^{(1)}  =
\left(
\begin {array}{ccc}
a_R &0& 0\\
0& 0& 0\\
0&0&0
\end {array}
\right) , h^{(2)}  =
\left(
\begin {array}{ccc}
0 &0& 0\\
0& c_R& d_R\\
0&d_R&c_R
\end {array}
\right), &
 \eea
Thus when the Higgs singlets $\Delta$ acquire vevs $\left(\D_1^0, \D_2^0\right)$ we get the following form for $M_R$,
\bea
M_R & = &
\left(
\begin {array}{ccc}
\D_1^0\, a_R &0& 0\\
0& \D_2^0\, c_R& \D_2^0\, d_R\\
0&\D_2^0\, d_R &\D_2^0\, c_R
\end {array}
\right).
\eea
\end{itemize} which of the form of Eq.(\ref{formM_R}) with $B_R=0$.
The analysis of the last subsection shows then that the deformation $\a$ in $M_D$ resurfaces as a `sole' perturbation $\chi$ in $M_\n$ which would get the desired form of Eq.(\ref{pert-texture}) with $\chi$ given by Eq.(\ref{pertparameter}) after putting $B_R=E_D=0$:
\be
\chi = \frac{\alpha\,\left(d_R - c_R\right)\, \left(C_D + D_D\right)}{\left(D_D - C_D\right)\left(c_R + d_R\right) + \a \left(c_R D_D -d_R C_D\right)}.
\ee

Before ending this section, we would like to mention that having multiple Higgs doublets in our constructions might display flavor-changing neutral currents. However, the effects are calculable and in principle one can adjust the Yukawa couplings so that to suppress processes like $\m \rightarrow e \g$ \cite{hagedorn}. Moreover, the constructions are carried out at the seesaw high scale, but the RG running effects are expected to be small when multiple Higgs doublets are present, and so we expect the predictions of the symmetry will still be valid at low scale.

\section{Discussion and summary}
We studied the properties of the $Z_2$ symmetry behind the $\mu-\tau$ neutrino universality. We singled out the texture ($S_+$) which imposes naturally a maximal atmospheric mixing
$\t_{23}=\pi/4$ and vanishing $\t_{13}$. The remaining mixing angle $\t_{12}$ remains free, and the other $Z_2$ necessary to characterize the neutrino mass matrix can be
used to fix it at its experimentally measured value ($\sim 33^0$). We showed how the $S_+$-texture accommodates all the neutrino mass hierarchies. Later, we implemented the $S_+$-symmetry in the whole lepton sector, and showed how it can accommodate the charged lepton mass hierarchies with small mixing angles of order of the `acute' charged lepton
mass hierarchies. We computed, within type-I seesaw, the CP asymmetry generated by the symmetry and found that the phases of the RH Majorana fields may be adjusted to produce enough baryon asymmetry. The fact that the $\mu$--$\tau$ symmetry does not determine fully the mixing angles, but leaves $\t_{12}$ as a free parameter able to take different values
 in $M_R$ and $M_D$ is crucial for obtaining leptogenesis within type-I seesaw scenarios.  We found also that ``complex-valued" perturbations on Dirac neutrino mass matrix can account for the correct neutrino mixing angles.

 We carried out a complete numerical study to find phenomenologically acceptable $M_\n$ respecting the approximate $S_+$, and we generated possible corresponding $M_R$ and $M^D_\n$.
 Crucially, we found in our numerical scanning that no ``real-valued" neutrino mass matrices can account for the experimental constraints, and so one has to take complex matrices from
 the outset. The perturbation at the level of $M_\n$ should also be complex in order to account for phenomenology. .

   Finally, we presented a theoretical realization of the perturbed Dirac mass matrix, where the
symmetry is broken spontaneously and the perturbation parameter originates from ratios of different Higgs fields vevs.
%%%%%%%%%%%%%%%%%%%%%%%%%%%%%%%%%%%%%

\section*{{\large \bf Acknowledgements}}
Part of the work was done
within the short visits program of ICTP.\\  N.C. thanks Graham Ross for discussions, and
acknowledges funding provided by the Alexander von Humboldt Foundation.

\end{document}